\documentclass[aps,prx,twocolumn,superscriptaddress,showpacs,floatfix,longbibliography]{revtex4-1}
\usepackage{amsmath,amssymb,amsfonts,float,graphics,epsfig,color,verbatim,tabularx,bm,multirow,appendix}

\usepackage{xcolor}
\usepackage{CJK}

\IfFileExists{newtxtext.sty}
   {\usepackage{newtxtext,newtxmath}}
   {\IfFileExists{stix.sty}
      {\usepackage{stix}}
      {\IfFileExists{mathptmx.sty}
      {\usepackage{mathptmx}}{} } }

\usepackage{textcomp}
\usepackage{bm}

\IfFileExists{siunitx.sty}{\usepackage{booktabs,siunitx}}{}

\definecolor{LinkColor}{rgb}{0.256,0.439,0.588}
\usepackage{hyperref}

\renewcommand{\vec}[1]{\mathbf{#1}}

\usepackage{pifont}

\newcommand{\pay}[1]{\textcolor{black}{#1}}

\begin{document}

\begin{CJK*}{UTF8}{gbsn}

\title{Solving quantum rotor model with different Monte Carlo techniques}

\author{Weilun Jiang (姜伟伦)}
\affiliation{Beijing National Laboratory for Condensed Matter Physics and Institute of Physics, Chinese Academy of Sciences, Beijing 100190, China}
\affiliation{School of Physical Sciences, University of Chinese Academy of Sciences, Beijing 100190, China}
\author{Gaopei Pan (潘高培)}
\affiliation{Beijing National Laboratory for Condensed Matter Physics and Institute of Physics, Chinese Academy of Sciences, Beijing 100190, China}
\affiliation{School of Physical Sciences, University of Chinese Academy of Sciences, Beijing 100190, China}
\author{Yuzhi Liu (刘毓智)}
\affiliation{Beijing National Laboratory for Condensed Matter Physics and Institute of Physics, Chinese Academy of Sciences, Beijing 100190, China}
\affiliation{School of Physical Sciences, University of Chinese Academy of Sciences, Beijing 100190, China}
\author{Zi Yang Meng (孟子杨)}
\affiliation{Department of Physics and HKU-UCAS Joint Institute of Theoretical and Computational Physics,
The University of Hong Kong, Pokfulam Road, Hong Kong, China}
\affiliation{Beijing National Laboratory for Condensed Matter Physics and Institute of Physics, Chinese Academy of Sciences, Beijing 100190, China}
\affiliation{Songshan Lake Materials Laboratory, Dongguan, Guangdong 523808, China}

\begin{abstract}
We systematically test the performance of several Monte Carlo update schemes for the $(2+1)$d XY phase transition of quantum rotor model. By comparing the local Metropolis (LM), LM plus over-relaxation (OR), Wolff-cluster (WC), hybrid Monte Carlo (HM), hybrid Monte Carlo with Fourier acceleration (FA) scheme, it is clear that among the five different update schemes, at the quantum critical point, the WC and FA schemes acquire the smallest autocorrelation time and cost the least amount of CPU hours in achieving the same level of relative error, and FA enjoys a further advantage of easily implementable for more complicated interactions such as the long-range ones. These results bestow one with the necessary knowledge of extending the quantum rotor model, which plays the role of ferromagnetic/antiferromagnetic critical bosons or Z$_2$ topological order, to more realistic and yet challenging models such as Fermi surface Yukawa-coupled to quantum rotor models. 
\end{abstract}



\maketitle

\end{CJK*}

\section{Introduction}
\label{sec:i}
The study of the critical behavior in XY model dates back to the early stage of renormalization group~\cite{Jose1977}. To date, very accurate analytical and numerical calculation at the $(2+1)$d O(2) Wilson-Fisher quantum critical point exist with high precision of exponents determined~\cite{Wallin1994,Hasenbusch1999,Hasenbusch2019,Lan2012,WanwanXu2019,Campostrini2001,Campostrini2006,Chester2019}, and the rich physics of such transition related with the superconductor-insulator~\cite{Fisher1988,Cha1991}, superfluid-insulator~\cite{Fisher1989,Greiner2002} and easy-plane quantum magnetic~\cite{Meng2008} transitions have been well acknowledged by the community. Moreover, the presence of Kosterlitz-Thouless (KT) transition at finite temperature also illustrates the nontrivial topological character of the setting and the associated vortex excitations are appearing in various material realizations~\cite{HanLi2019,ZeHu2020,YDLiao2021}. In short, the quantum XY criticality and KT physics originated from the $(2+1)$d O(2) transition are rich and profound.

With the acknowledgement of its importance, the renormalization group expansion calculations have been performed upon the $(2+1)$d O(2) Wilson-Fisher fixed point and comparison with the unbiased Monte Carlo simulation results are achieved~\cite{Campostrini2001,Campostrini2006}, lately conformal bootstrap calculation has also been succeeded in $(2+1)$d O(2) QCP~\cite{Chester2019}. Among different Monte Carlo simulation methods, such as local Metropolis~\cite{Wallin1994}, Swendsen-Wang and Wolff-cluster~\cite{SwendsenWang1987,Wolff1989}, over-relaxation~\cite{Adler1981}, worm-algorithm in the path-integral formalism~\cite{WanwanXu2019,Alet2003} etc, accurate results have also been obtained. The remaining issue is that there still lacks systematic analysis and comparison of the performance of various Monte Carlo schemes, both in terms of the autocorrelation time and physical CPU hours in achieving the same level of numerical accuracy. In this work, we would like to fill in this gap.

We implement the Monte Carlo simulation for $(2+1)$d quantum rotor model and focus on the performance of simulation in the vicinity of the $(2+1)$d XY quantum critical point. Among the five different update schemes we tested in this work, which are comprised of local Metropolis (LM), LM plus over-relaxation (OR), Wolff-cluster (WC), hybrid Monte Carlo (HM), hybrid Monte Carlo with Fourier acceleration (FA), we find that to achieve the same level of numerical accuracy of the physics observables at the $(2+1)$d O(2) QCP, the WC and FA schemes have the smallest autocorrelation time and cost the least amount of CPU hours. Moreover, since FA scheme is more versatile in terms of implementation for complicated Hamiltonians, it has the advantage towards the future development of the quantum many-body simulations in which the simulation of O(2) lattice boson is the central ingredient. 

For example, the Fermi surface Yukawa-coupled to critical O(2) bosons at $(2+1)$d will be the natural extension of the system of Fermi surface Yukawa-coupled to critical Ising bosons where concrete numerical evidence of the non-Fermi-liquid~\cite{XYXu2017,ZHLiu2019} and quantum critical scaling beyond the Hertz-Mills-Moriya framework~\cite{ZHLiu2019} have been revealed recently. Also, when the gauge field with U(1) symmetry couples to matter field at $(2+1)$d, such as the Dirac fermion in the recent case~\cite{XYXu2019,WangWei2019}, although attempt succeeded in reaching small to large system sizes and established the existence of the U(1) deconfined phase even if the fermion flavor is just $N_f=2$~\cite{XYXu2019}, larger system sizes are inevitably crucial to further confirm such important discovery at the thermodynamic limit. Of course the Monte Carlo simulation of these systems involves the update of fermionic determinant and therefore their computational complexity also comes from the fermionic part of the configurational weight~\cite{XYXuReview2019}, the overall bottleneck can be overcome by finding efficient update scheme of the U(1) (O(2)) gauge (boson) fields on the lattice upon the low-energy effective model extracted from methods such as the self-learning approach~\cite{JWLiu2017,XYXuSelf2017,HYLu2021}. These effective models usually acquire long-range interactions beyond the bare bosonic ones in the original Hamiltonian, and are difficult to handle with conventional methods such LM and WC discussed here. 


Also, the XY and KT-related physics appear in several recently discovered frustrated magnets, such as the compound of TmMgGaO$_4$, which nicely develops a Kosterlitz-Thouless melting of magnetic order in a triangular quantum Ising model setting~\cite{HanLi2019,YDLiao2021,ZeHu2020}. One could certainly envision the application of different Monte Carlo schemes tested here to perform better simulation upon future models of quantum XY magnets. 

\pay{We end the introduction by briefly outline the structure of the paper.} In Sec.~\ref{sec:ii}, we describe the quantum rotor model and setup its path-integral formulation upon which the quantum Monte Carlo simulation will be carried out. In Sec.~\ref{sec:iii}, we explain the different Monte Carlo update schemes employed in this work, and pay more attention to the HM and FA schemes which are less used in a condensed matter setting. Then Sec.~\ref{sec:iv} offers the results and compares the autocorrelation time and CPU hours of these methods at the $(2+1)$d O(2) QCP in details. Sec.~\ref{sec:v} summarizes the main results and elaborates more on the relevance of this work towards to more frontier models in which the successful simulation of the quantum rotor model is of vital importance. 


\begin{figure}[htp!]
	\includegraphics[width=\columnwidth]{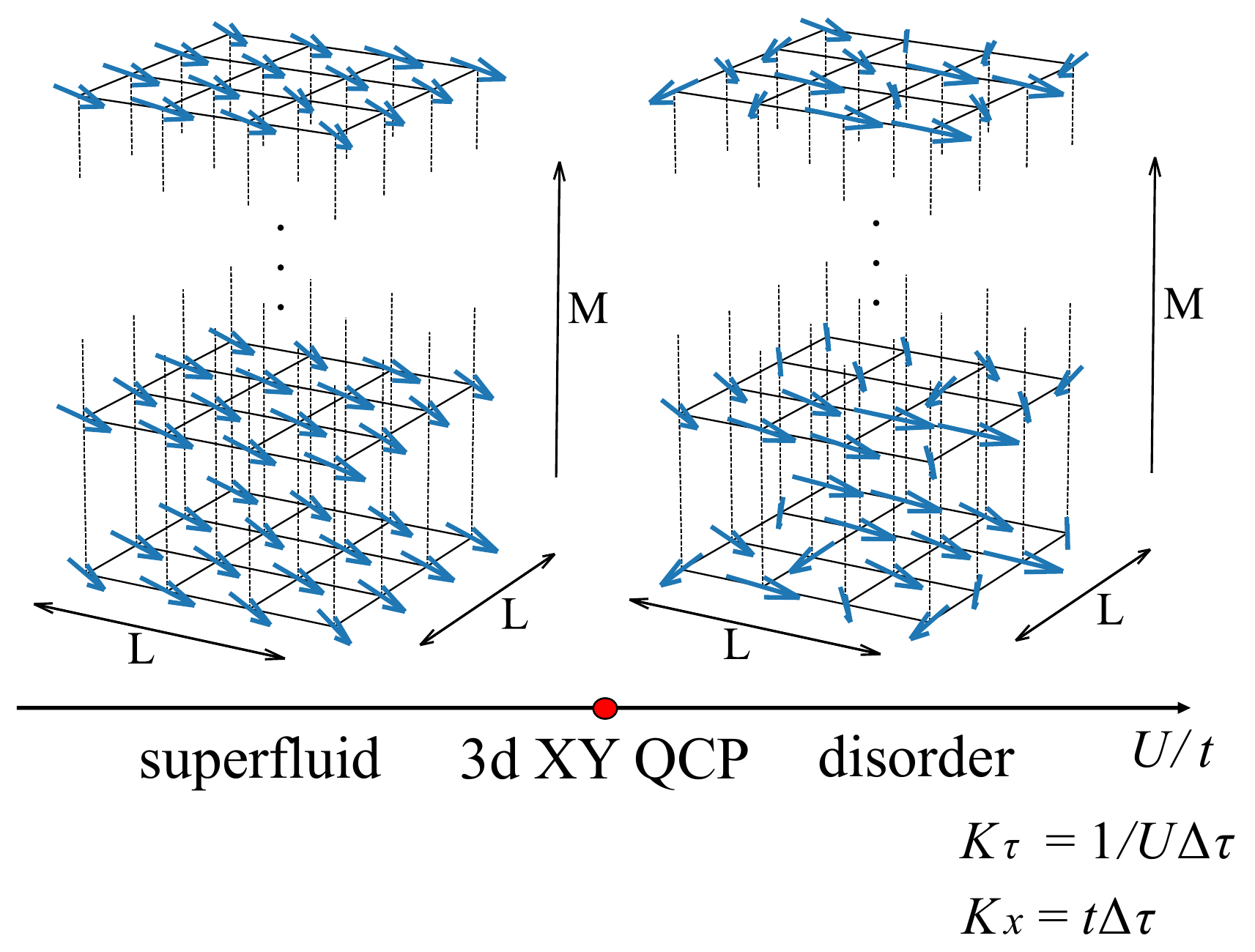}
	\caption{ Schematic plot of the configuration space for the $(2+1)$d O(2) XY QCP. The blue arrows in the space-time coordinate stand for the unit vector $\{\theta_{i}(l)\}$ (or $\{\theta_{\mathbf{r}}\}$ with $\mathbf{r}$ the space-time coordinate) in our simulation and the $K_\tau$ and $K_x$ are the anisotropic temporal and spatial interaction strengths in the path-integral of Eq.~\eqref{eq:eq15}.  Along the axis of $\frac{U}{t}$, quantum critical point $g_c=(\frac{U}{t})_c$ separates the superfluid phase with O(2) XY symmetry-breaking and the disordered phase where the system is in a trivial and symmetric state.  }
	\label{fig:fig1}
\end{figure}

\section{Model}
\label{sec:ii}
We begin the discussion with the 2d Bose-Hubbard model on square lattice~\cite{Fisher1988,Wallin1994}, 
\begin{equation}
H_{\text{BH}} = -t\sum_{\langle i,j \rangle} (b_{i}^{\dagger} b_{j} + b_{j}^{\dagger} b_{i})  
+\frac{U}{2} \sum_{i} n_{i}( n_{i}-1) + \mu\sum_{i} n_{i},
\label{eq:eq1}
\end{equation}
where $t$ is the nearest-neighbor hopping strength of boson, $\mu$ is the chemical potential, and $U$ is the on-site repulsion. The creation and annihilation operator of boson satisfy commutation relation $[b_{i},b^{\dagger}_{j}]=\delta_{i,j}$ where $\delta$ is the Kronecker function. At a fixed chemical potential, Eq.~\eqref{eq:eq1} describes the quantum phase transition from superfluid to Mott insulator as a function of $\frac{U}{t}$. If the average filling of boson is an integer, then the transition is of $(2+1)$d O(2) universality with the dynamical exponent $z=1$, and if the average boson filling is deviated from integer, the transition is of $(2+2)$d O(2) universality with the dynamical exponent $z=2$~\cite{Fisher1988,Fisher1989}. In the former case, one can write $b_{i}$ as $|b_{i}|e^{i\theta_i}$ and integrate out the amplitude fluctuations. Then the BH model becomes a model of coupled of Josephson junctions~\cite{Fisher1989}, 
\begin{equation}
H_{\text{JJ}} = \frac{U}{2}\sum_i {n_i}^2  - t\sum_{ \langle i,j \rangle }\cos (\theta_i - \theta_j).
\label{eq:eq2}
\end{equation}
In this form, $n_{i}$ runs from $-\infty$ to $+\infty$ of integers. The commutation relation of the bosonic operators in Eq.~\eqref{eq:eq1} now translates to the commutation relation between $\theta$ and $n$,
\begin{equation}
[\theta_i , n_j] = i\delta_{i,j}.
\end{equation}
This means that one can write the Hamiltonian Eq.~\eqref{eq:eq2} in an coherent state representation of the angle $\theta$, and the Hamiltonian can be written in the form of quantum rotor model~\cite{Cha1991,Fisher1988}, 
\begin{equation}
H_{\text{qr}} = \frac{U}{2}\sum_i \big(-i\frac{\partial}{\partial \theta _i}\big)^2  - t\sum_{ \langle i,j \rangle } \cos (\theta _i - \theta_j).
\label{eq:eq4}
\end{equation}

The derivative $\frac{\partial}{\partial \theta}$ plays the role of angular momentum and can be further expressed in a path integral of the coherent state such that its Monte Carlo simulation becomes possible. We illustrative this process starting from the partition function,
\begin{equation}
   \begin{aligned}
Z &= \text{Tr} \exp[-\beta(T+V)] \\    &= \text{Tr}\exp[-\beta\big(-\frac{U}{2}\sum_i\frac{\partial^2}{\partial\theta^{2}_{i}}-t\sum_{\langle i,j \rangle}\cos(\theta_i - \theta_j)\big)]  
   \end{aligned}
\end{equation}
where $T$ represents the kinetic energy and $V$ the interaction energy. Using this shorthand notation, one can Trotter the path integral as,
\begin{equation}
   \begin{aligned}
Z &= \text{Tr}\{\exp[-\beta(T+V)]/M\}^{M} \\
  &= \lim_{M\to\infty}\text{Tr}\{\prod_{l=0}^{M-1}\exp[-\Delta\tau T]\exp[-\Delta\tau V]\}
  \label{eq:eq6}
   \end{aligned}
\end{equation}
where the imaginary time $\beta$ has been divide into $M$ slices with step $\Delta\tau=\frac{\beta}{M}$ and we index the time slices with label $l\in[0,M-1]$. Now one can insert the complete sets of the coherent state of $\{\theta(l)\}$ at each imaginary time step in Eq.~\eqref{eq:eq6} and have,
\begin{equation}
Z = \int \mathcal{D}\theta \prod_{l=0}^{M-1} \langle\{\theta(l+1)\}|\exp[-\Delta\tau T] \exp[-\Delta\tau V]|\{\theta(l)\}\rangle.
\label{eq:eq7}
\end{equation}
It is clear at this step that $\theta_i(l)$ spans the space-time configuration space of $L\times L\times M$ that we will use Monte Carlo to sample, and the states should follow the periodic boundry condition $\{\theta(M)\}=\{\theta(0)\}$. Such a setting of configuration space is shown in Fig.~\ref{fig:fig1}.

Now we can look into the detailed form of $V$ and $T$. For the potential term, the coherent state is its eigenstate, thus, it becomes,

\begin{equation}
\exp[-\Delta\tau V]|\{\theta(l)\}\rangle = \exp\big\{\Delta\tau t \sum_{\langle i,j \rangle}\cos[\theta_i(l)-\theta_j(l)]\big\}|\{\theta(l)\}\rangle
\end{equation}
and consequently the partition function in Eq.~\eqref{eq:eq7} becomes,
\begin{equation}
Z\approx\int \mathcal{D}\theta \prod_{l=0}^{M-1}\exp\big\{K_x \sum_{\langle i,j \rangle}\cos[\theta_i(l)-\theta_j(l)]\big\}T_l,
\end{equation}
where $K_x=t\Delta\tau$ and the remaining kinetic part is $T_l \equiv \langle \{\theta(l+1)\} | e^{-\Delta\tau T}|\{\theta(l)\}\rangle$. For the kinetic term, different sites commute with each other, so it can be written in the form of products over the spatial lattice of $i\in L\times L$,  
\begin{equation}
T_l = \prod_{i} \langle \theta_{i}(l+1)|\exp\big[\frac{\Delta\tau U}{2}\frac{\partial^{2}}{\partial\theta_i(l)^2}\big]|\theta_i(l)\rangle.
\label{eq:eq10}
\end{equation}
If one denotes $J_i(l)$ the integer-valued angular momentum at site $i$ and time $l$, then one has $\langle \theta_i(l)|J_i(l)\rangle = e^{iJ_{i}(l)\theta_i(l)}$ as the eigenfunction of the kinetic energy operator, then Eq.~\eqref{eq:eq10} becomes,
\begin{widetext}
\begin{equation}
T_l = \sum_{\{J\}}\prod_{i} \langle \theta_i(l+1)|J_i(l)\rangle \exp\big\{-\frac{\Delta\tau U}{2}[J_i(l)]^2\big\}\langle J_i(l)|\theta_i(l)\rangle,
\end{equation}
and eventually the partition function becomes,
\begin{equation}
Z \approx \int\mathcal{D}\theta \sum_{\{J\}} \exp\big\{K_x \sum_{\langle i,j \rangle}\sum_{l=0}^{M-1}\cos[\theta_i(l)-\theta_j(l)]\big\}\exp\big\{-\frac{\Delta\tau U}{2} \sum_{i}\sum_{l=0}^{M-1}[J_i(l)]^2\big\}\exp\big\{i\sum_{i}\sum_{l=0}^{M-1}J_i(l)[\theta_i(l+1)-\theta_i(l)]\big\},
\label{eq:eq12}
\end{equation}
where the configuration space is spanned by the product of  $\{\theta_i(l)\}$ and $\{J_i(l)\}$ with $i\in L\times L$ and $l\in M$.
\end{widetext}
 
From here on one can have two ways to simulate Eq.~\eqref{eq:eq12}. One is by integrating out the variable $\{\theta_{i}(l)\}$ and arrives at a link model with integer-valued $\{J_{i}(l)\}$ on every bond. This type of algorithm is not the main focus of this paper and we discuss it in the Appendix~\ref{app:appA}.

In this paper, however, we choose the other way to simulate partition function Eq.~\eqref{eq:eq12} by summing over the variable $\{J_{i}(l)\}$. In doing so, we need to first use the Poisson summation and Gaussian integral, to change the last two terms in Eq.~\eqref{eq:eq13} and obtain,
\begin{equation}
   \begin{aligned}
F(\theta) \equiv \sum_{J}e^{-\frac{\Delta\tau U}{2} J^2}e^{iJ\theta} &= \sum_{m=-\infty}^{\infty}\int_{-\infty}^{\infty}dJ e^{2\pi i J m}e^{-\frac{\Delta\tau U}{2}J^2}e^{iJ\theta} \\
&= \sum_{m=-\infty}^{\infty}\sqrt{\frac{2\pi}{\Delta\tau U}} e^{-\frac{1}{2\Delta\tau U}(\theta - 2\pi m)^2}.
\label{eq:eq13}
   \end{aligned}
\end{equation}
Since $\Delta\tau$ is small, the summation over of $J$ slowly convergences. Then we perform Villian approximation\cite{Fisher1989_2} on Eq.~\eqref{eq:eq14},
\begin{equation}
F(\theta) \approx e^{K_\tau \cos(\theta)},
\label{eq:eq14}
\end{equation}
where $K_\tau = \frac{1}{U\Delta\tau}$. Combining the cosine functions along the imaginary time axis and the spatial axis, the partition function of Eq.~\eqref{eq:eq12} arrives at a 3d anisotropy classical XY model,
\begin{equation}
Z = \int\mathcal{D}\theta \exp\big\{\sum_{\langle \mathbf{r},\mathbf{r'}\rangle} K_{\langle \mathbf{r},\mathbf{r'}\rangle}\cos(\theta_{\mathbf{r}}-\theta_{\mathbf{r'}})\big\},
\label{eq:eq15}
\end{equation}
where $K_x = t\Delta\tau$, the summation is now over nearest-neigbhor bonds in both space and time directions, i.e. $\mathbf{r}=(i,j,l)$. At the limit of $\Delta\tau \rightarrow 0$, the interactions become more and more anisotropic as $K_x \rightarrow 0$ and $K_\tau \rightarrow \infty$. However, it is easy to see that their geometric mean $K=(K_x K_\tau)^{\frac{1}{2}}=\frac{t}{U}$ is kept finite and it is the control parameter of the $(2+1)$d O(2) transition. The configuration space $\{\theta_{i}(l)\}$, the two phases separated by the $(2+1)$d O(2) QCP and the $K_x$ and $K_\tau$ couplings are depicted in Fig.~\ref{fig:fig1}.

\section{Algorithm}
\label{sec:iii}
The quantum rotor model in Eqs.~\eqref{eq:eq2},\eqref{eq:eq12} and \eqref{eq:eq15} can be solved with various Monte Carlo simulation schemes, including the local Metropolis (LM)~\cite{Metropolis1953}, LM plus over-relaxation (OR)~\cite{Adler1981,Lan2012}, Wolff-cluster (WC)~\cite{Wolff1989}, hybrid Monte Carlo (HM) and hybrid Monte Carlo supplemented with Fourier acceleration (FA)~\cite{Davies1986}. In this section, we will elucidate the basic steps in these schemes with the detailed explanation in HM~\cite{Ferreira1993,Duane1988Hybrid,Simon2002Testing} and FA~\cite{Batrouni1985} schemes as they are more used in the high-energy community and less so to condensed matter.

\subsection{Local update}
This is the standard Monte Carlo update scheme, based on the Metropolis-Hastings algorithm~\cite{Metropolis1953,Hastings1970}. In the rotor model setting, the update is comprised of the following steps: first we randomly choose one site, and then try to change the $\theta_i(l)$ by a random value within 0 to 2$\pi$. The acceptance of such update is determined by the Metropolis acceptance ratio. To satisfy ergodicity, we choose the site with its lattice index in sequence and define one Monte Carlo update step as a sweeping over the entire space-time configuration when calculating the autocorrelation time.

\subsection{Over-relaxation scheme}
Over-relaxation method~\cite{Adler1981} is an improvement of the local update. It was introduced in Monte Carlo evaluation of the partition function for multiquadratic actions~\cite{Whitmer1984} and have been used in quantum rotor model~\cite{Lan2012}. For each site, the $\theta_{\mathbf{r}}$ can be mapped to one unit vector $\Theta_{\mathbf{r}}$ with its angle of rotation between 0 to $2\pi$. The method regards the total effect of six nearest sites (four spatial and two temporal) of $\mathbf{r}$ as a new vector field, $\mathbf{H}_\mathbf{r}$. Where $\mathbf{H}_\mathbf{r}=\sum_{\langle \mathbf{r},\mathbf{r'}\rangle} \Theta_{\mathbf{r'}}$. One can design an update from $\Theta_{\mathbf{r}}$ to $\Theta'_{\mathbf{r}}$ such that the energy between this site and its neighboring sites, i.e., the vector dot product of $\Theta_{\mathbf{r}}\cdot \mathbf{H_r}$ and $\Theta'_{\mathbf{r}}\cdot \mathbf{H_r}$, is conserved. One can then easily write down the following relation,
\begin{equation}
\Theta'_{\mathbf{r}} =  - \Theta_{\mathbf{r}} + 2\frac{{{\Theta_{\mathbf{r}}} \cdot {\mathbf{H}_\mathbf{r}}}}{{{|\mathbf{H}_\mathbf{r}|}^2}}{\mathbf{H}_\mathbf{r}}.
\label{eq:eq16}
\end{equation}
Therefore, the $\Theta'_{\mathbf{r}}$ has the same energy with its neighbors as that of $\Theta_{\mathbf{r}}$. Thus the update can be accepted with $100\%$ certainty, and guarantees the best acceptance rate. However, since the over-relaxation scheme strictly conserves the energy among different configurations, it will not be ergodic and has to be supplemented together with other update scheme such as the LM.  In this paper, we use one local update sweep and one over-relaxation sweep as one update sweep for the OR scheme. As shown in Sec.~\ref{sec:ivb}, that OR scheme is indeed faster than the LM scheme.

\subsection{Cluster update}
Here we employ the Wolff update scheme, it is one of the effective cluster update method~\cite{Wolff1989}. Since our model is XY model with O(2) symmetry, we can easily construct the global Wolff cluster. The basic principle is to grow a cluster with certain probability and change all of the site in the cluster. For our model, we choose a random site at first and grow the cluster with the probability,
\begin{equation}
   \begin{aligned}
   P({\Theta _\mathbf{r}},{\Theta _\mathbf{r'}}) &= 1 - \exp \{ \min [ 0, -K_{\langle \mathbf{r},\mathbf{r'}\rangle} {\Theta _\mathbf{r}} \cdot (1 - R(\hat r)) \Theta _\mathbf{r'}] \}  \\
   &= 1 - \exp \{ \min [ 0,-2 K_{\langle \mathbf{r},\mathbf{r'}\rangle} (\hat r \cdot {\Theta _\mathbf{r}})(\hat r \cdot \Theta _\mathbf{r'}) ] \} ,
\label{eq:eq17}
   \end{aligned}
\end{equation}
where $\Theta_{\mathbf{r}}$ are the neighboring unit vectors as before and $\hat r$ stands for a random unit vector pointing towards a direction within the angle of 0 to 2$\pi$. We define $R(\hat r)$ to be the operation of mirror symmetry along the mirror direction normal to $\hat r$.
In one Monte Carlo update, we randomly choose one site and one vector $\hat r$ and grow cluster in both spatial and temporal bonds, then reorient all the $\theta_\mathbf{r}$ in the cluster with respect to the mirror direction normal to $\hat r$. The sketch map of detailed update process is shown in Appendix~\ref{app:appB}. As will be shown in Sec.~\ref{sec:ivb}, the cluster update has much smaller autocorrelation time than the local and over-relaxation update schemes.

\subsection{Hybrid Monte Carlo}
Hybrid Monte Carlo~\cite{DUANE1987216,Gupta1998} is widely used in high-energy physics to simulate the equilibrium distribution of many-particle system. The original form of it is the real time-evolution of classical system for classical Hamiltonian. It has also been used in the condensed matter system to carry on molecular dynamics simulation~\cite{Mehlig1992}. In quantum Monte Carlo, the time-evolution can generate canonical distribution and offers a way to produce sample space of the Markov chain. There are attempts to implement it to simulate correlated electron systems, for example in a Hubbard model setting~\cite{Scalettar1987,Beyl2018}.

Here we discuss the basic steps of HM scheme, and focus on its quantum rotor model implementation. 

First, we add one auxiliary parameter $p_{\mathbf{r}}$ to every site in the configuration space and extend the partition function in Eq.~\eqref{eq:eq15} to,
\begin{equation}
   \begin{aligned}
   Z &=Z \times 1 \\
   &= Z \left( C \int \mathcal{D}p \exp (-\beta\frac{p_{\mathbf{r}}^2}{2m_{\mathbf{r}}}) \right)\\
   &=C \int \mathcal{D}p \int \mathcal{D}\theta \exp (-\beta\frac{p_{\mathbf{r}}^2}{2m_{\mathbf{r}}}) \exp (-\beta H_{\text{qr}}(\theta_{\mathbf{r}})),
   \end{aligned}
   \label{eq:eq18}
\end{equation}
where $\int\mathcal{D}p=\prod_\mathbf{r} dp_{\mathbf{r}}$, $C$ is a normalization coefficient and $H_{\text{qr}}$ is the original quantum rotor Hamiltonian in Eqs.~\eqref{eq:eq2}, \eqref{eq:eq12} and \eqref{eq:eq15}. Now the configuration space is extend to $\{\theta_{\mathbf{r}},p_{\mathbf{r}}\}$, with $p_{\mathbf{r}}$ serving as the canonical momentum of the canonical coordinate $\theta_{\mathbf{r}}$. Thus, the Hamiltonian can now be written as,
\begin{equation}
   \begin{aligned}
   H &= K(p) + H_{\text{qr}}(\theta) = \frac{{{P^2}}}{{2m}} + {H_{\text{qr}}}(\theta) \\
   &= \sum_{\mathbf{r}}\frac{{{p_\mathbf{r}^2}}}{{2m}} - \sum_{\langle \mathbf{r},\mathbf{r'}\rangle} \frac{K_{\langle \mathbf{r},\mathbf{r'}\rangle}}{\beta}\cos(\theta_{\mathbf{r}}-\theta_{\mathbf{r'}}).
   \end{aligned}
\label{eq:eq19}
\end{equation}
Here, $K(p)$ is the kinetic energy and $H_{\text{qr}}(\theta)$ becomes the potential energy term in hybrid Hamiltonian of Eq.~\eqref{eq:eq19}. The spirit of hybrid Monte Carlo is to use the Hamiltonian dynamics,
\begin{equation}
   \begin{aligned}
      \frac{d\theta_{\mathbf{r}}}{dt} &= \frac{\partial H}{\partial p_\mathbf{r}}\\
      \frac{dp_{\mathbf{r}}}{dt} &= -\frac{\partial H}{\partial \theta_\mathbf{r}}
   \end{aligned}
\label{eq:eq20}
\end{equation}
to generate a new configuration $\{\theta_{\mathbf{r}},p_{\mathbf{r}}\}$ from time $t$ to time $t+\epsilon$ with $\epsilon$-the footstep in time evolution of Hamiltonian dynamics. In the computation, one needs to solve the differential equation to perform the time evolution. If the time interval $\epsilon$ is small enough, simple process like Euler method can propagate the system from the initial point at time $t$ to $t+\epsilon$,
\begin{equation}
   \begin{aligned}
   {p_\mathbf{r}}(t + \epsilon ) &= p_\mathbf{r}(t) + \epsilon \frac{{d{p_\mathbf{r}(t)}}}{{dt}} = {p_\mathbf{r}}(t) - \epsilon \frac{{\partial H_{\text{qr}}(\theta(t))}}{{\partial {\theta_\mathbf{r}}}}+O(\epsilon), \\ 
   {\theta_\mathbf{r}}(t + \epsilon ) &= {\theta_\mathbf{r}}(t) + \epsilon \frac{{d{\theta_\mathbf{r}(t)}}}{{dt}} = {\theta_\mathbf{r}}(t) + \epsilon \frac{p_\mathbf{r}(t)}{m_{\mathbf{r}}}+O(\epsilon),
   \end{aligned}
\label{eq:eq21}
\end{equation}
the systematic error is of the order $O(\epsilon)$ in the Euler method. The Hamiltonian evolution guarantees that the update is moving along the isoenergic surface, at least in principle, and the small energy difference will natually give rise to high acceptance ratio of the updated configuration. This is one of the advantage of hybrid Monte Carlo. 

If we conduct $N_{\text{HM}}$ times of the evolution, i.e., $t+N_{\text{HM}}\epsilon$, the trajectory in the phase space will move a long distance. Such substantial update of the configuration  $\{\theta_{\mathbf{r}},p_{\mathbf{r}}\}$ can be viewed as effectively global update, which could in principle reduce the autocorrelation time at the QCP. One point we need to pay attention to is that the detailed balance condition requests the evolution in configuration space to respect the time-reversal symmetry and actually the Euler method does not satisfy this condition. So in the real simulation we use leapfrog method of Hamiltonian dynamics, 
\begin{equation}
   \begin{aligned}
      {p_\mathbf{r}}(t + \epsilon /2) &= {p_\mathbf{r}}(t) - (\epsilon /2)\frac{{\partial H_{\text{qr}}(\theta(t))}}{{\partial {\theta_\mathbf{r}}}}+O(\epsilon^2), \\
      {\theta_\mathbf{r}}(t + \epsilon ) &= {\theta_\mathbf{r}}(t) + \epsilon \frac{p_\mathbf{r}(t + \epsilon /2)}{m_{\mathbf{r}}}+O(\epsilon^2),\\
      {p_\mathbf{r}}(t + \epsilon ) &= {p_\mathbf{r}}(t + \epsilon /2) - (\epsilon /2)\frac{{\partial H_{\text{qr}}(\theta(t + \epsilon ))}}{{\partial {\theta_\mathbf{r}}}}+O(\epsilon^2),
   \end{aligned}
\label{eq:eq22}
\end{equation} 
with the systematical error of $O(\epsilon^{2})$.

Finally, the acceptance ratio of the $\{\theta_\mathbf{r},p_\mathbf{r}\}$ configuration after the $N_{\text{HM}}$ steps time evolution can be evalued with respect to the hybrid Hamiltonian in Eq.~\eqref{eq:eq19} is,
\begin{equation}
P_{\text{acc}} = \min \{ 1,\exp [ - \beta(H(\{\theta',p'\}) - H(\{ \theta, p \}))]\}.
\label{eq:eq23}  
\end{equation} 
Overall, the $\{ p \}$ is an auxiliary degree of freedom to help to generate uncorrelated configuration in $\{\theta\}$ with high acceptance ratio. So after the acceptance of the update, one can easily regenerate a new $\{p\}$ configuration and start the next step of the time evolution of the hybrid Hamiltonian, and can evaluate the acceptance of such step once the $N_{\text{HM}}$ steps' time evolution is complete. This process is therefore the Markov chain for the hybrid Monte Carlo. To satisfy the detailed balance condition, each $p$ should be generated as,
\begin{equation}
   P(p) \propto \exp(-\beta \frac{p^2}{2})
   \label{eq:eq24}  
\end{equation} 
We summarize the steps of HM with the pseudo-code in Tab.~\ref{tab:tab1}

\begin{table}[htp!]
\caption{pseudocode of HM algorithm}
   \begin{tabular}{l}
\hline
1.Generate $\{p\}$ for each site by Gaussian distribution and 
obtain the \\
\quad configuration $\{\theta,p\}$. \\

2.Calculate the potential energy $H_{\text{qr}}$ and kinetic energy $K$ of the \\
\quad model in the hybrid Hamiltonian in Eq.~\eqref{eq:eq19}. \\

3.do $n=1$, $N_{\text{HM}}$ \\

\qquad $\{\theta(t),p(t)\}$ $\rightarrow$ $\{\theta(t+\epsilon),p(t+\epsilon)\}$ with Eq.~\eqref{eq:eq22}\\

\quad end do \\

4.Calculate the potential energy $H'_{\text{qr}}$ and kinetic energy $K'$ of the \\
\quad model. \\

5.Use Eq.~\eqref{eq:eq23} to determine whether the new configuration will be \\ 
\quad accepted. \\

6.Step 1-5 is one whole update. Further iterate step 1-5 to continue \\
\quad the Markov chain.\\
\hline
\label{tab:tab1}
   \end{tabular}
\end{table}

In our simulation, we define $m_{\mathbf{r}}=1$ and choose $\epsilon=0.3$ for system size $L=6$. These parameters should be tested before putting into production runs. For example, if $\epsilon$ is too small, each update only moves a short distance in the configuration space, still leads to high autocorrelation time. While if $\epsilon$ is too large, the energy difference will also be large, and cause small acceptance ratio. From the high energy HM literature~\cite{Gupta1998} -the optimized hybrid step size $\epsilon$ is proportion to the $V^{-\frac{1}{4}}$, where $V = L\times L\times M$ is the space-time volume of our configuration space. The choice of number of evolution times $N_{\text{HM}}$ will directly determine the actual computation time, and will also give rise to high autocorrelation time if it is too small. In our simulation, we choose $N_{\text{HM}}=20$ and $\epsilon$ is decided by the $V^{-\frac{1}{4}}$ empirical rule. 

Although HM provides an effective non-local update scheme of the original configuration space $\{\theta_{\mathbf{r}}\}$, as will be shown in Sec.~\ref{sec:ivb}, it still suffers from critical slowing down at the $(2+1)$d XY critical point. And to finally overcome it, we will discuss a better Monte Carlo update scheme for the quantum rotor model: the Hybrid + Fourier Acceleration method. 

\subsection{Hybrid + Fourier Acceleration}
Hybrid Monte Carlo + Fourier Acceleration (FA) is designed to conquer the critical slowing down in the HM scheme. It was firstly proposed to be combined with another molecular dynamic method -- Langevin equation and here we use it in the HM~\cite{Davies1986,Batrouni1985}. 

The analysis of FA~\cite{Batrouni2019} reveals that the critical slowing down comes from the fact that in the internal dynamics of Monte Carlo, i.e., at the critical point, long(short) wave length mode which takes longer(shorter) time to evolve. And typical update schemes do not respect such fact and use the same time step to evolve both modes, which are consequently ended with long autocorrelation time of the Monte Carlo dynamics. To avoid this problem, one can add different footstep to different modes to make them evolve at the same speeds, i.e., evolve according to the internal dispersion relation of the Hamiltonian. In this way, the autocorrelation time at the critical point can be reduced, as we are now updating the configuration with the intrinsic dynamics of the Hamiltonian.

To simplify the process, for the quantum rotor model at hand, we only consider 1D chain in the temporal direction as $\cos(\theta_{i}(l) - \theta_{i}(l'))$. Thus the force ($-\frac{\partial H_{\text{qr}}}{\partial \theta_{i}(l)}$) at $(i,l)$ in the hybrid time evolution is equal to $\sin( \theta_{i}(l) - \theta_{i}(l') )$. When the difference between $\theta_{i}(l)$ and $\theta_{i}(l')$ is small (it's certainly the case for our quantum rotor model along the imaginary time, since $K_\tau = \frac{1}{U\Delta\tau}$ is large at small $\Delta\tau$), this term can be approximated as $( \theta_{i}(l) - \theta_{i}(l'))$. 

In the leapfrog process in HM scheme in Eq.~\eqref{eq:eq22}, we take $\epsilon$ as the step size independent on the modes and from the discussion above, it is clear that to reduce the autocorrelation time, the proper coefficients shall depend on the evolution speed of each mode determined by the force. We can write the evolution speed for every site as,

\begin{equation}
   \begin{aligned}
   v(\theta_i(l)) &\equiv F(\theta_i(l)) = - \frac{{\partial H_\text{{qr}}(\{\theta\})}}{{\partial {\theta_i(l)}}} \\
   &= \theta_{i}(l)-\theta_{i}(l+1)+\theta_{i}(l)-\theta_{i}(l-1).
   \label{eq:eq25}
   \end{aligned}
\end{equation}  
We then perform Discrete Fourier Transformation (DFT) along the 1D temporal chain and obtain the footstep for $p$-evolution,
\begin{equation}
   \theta_{i}(l)=\frac{1}{M}\sum_{p_\tau} \theta_{i}(p_\tau) \exp(\frac{i2\pi l}{M} p_\tau )
   \label{eq:eq26}
\end{equation}
\begin{widetext}
where $p_\tau=0,1,2,\cdots,M-1$ and Eq.~\eqref{eq:eq25} then becomes,
\begin{equation}
   \begin{aligned}
   v(\theta_i(l)) &=\frac{1}{M} \sum_{p_\tau} [2\theta_{i}(p_\tau) \exp(\frac{i2\pi l}{M} p_\tau )-\theta_{i}(p_\tau) \exp(\frac{i2\pi (l+1)}{M} p_\tau ) -\theta_{i}(p_\tau) \exp(\frac{i2\pi (l-1)}{M} p_\tau )] \\
   &= \frac{1}{M}\sum_{p_\tau} [2-\exp(\frac{i2\pi}{M}p_\tau)-\exp(\frac{-i2\pi}{M}p_\tau)]\theta_{i}(p_\tau) \exp(\frac{i2\pi l}{M} p_\tau )\\
   &= \frac{1}{M} \sum_{p_\tau} [2-2\cos(\frac{2\pi p_\tau}{M})] \theta_{i}(p_\tau) \exp(\frac{i2\pi l}{\beta} p_\tau ) \\
    &= \frac{1}{M} \sum_{p_\tau} v(p_\tau) \exp(\frac{i2\pi l}{\beta} p_\tau ),
   \label{eq:eq27}
   \end{aligned}
\end{equation}
with,
\begin{equation}
  v_i(p_\tau) \equiv {F}_i(p_\tau)  = (2 - 2\cos (\frac{2\pi p_\tau}{M}))\theta_{i}(p_\tau).
\label{eq:eq28}
\end{equation}
\end{widetext}
We can observe that the small $p_\tau$ has slower speed of evolution. Now adding the reciprocal of this term makes long wave length mode evolve faster than before. Thus the leapfrog method with Fourier acceleration can be written as,
\begin{equation}
   \begin{aligned}
      {p_i(l)}(t + \epsilon /2) &= {p_i(l)}(t) - \frac{\epsilon }{2} {\mathbf{F}^{ - 1}}\omega (p_\tau)\mathbf{F}\frac{{\partial H_\text{qr}(\{\theta(t)\})}}{{\partial {\theta_{i}}}} \\
      {\theta_i(l)}(t + \epsilon ) &=  {\theta_i(l)}(t) + \epsilon {\mathbf{F}^{ - 1}}  \omega (p_\tau)\mathbf{F}{p_{i}}(t + \epsilon /2)\\
      {p_i(l)}(t + \epsilon ) &= {p_i(l)}(t + \epsilon /2) - \frac{\epsilon }{2}{\mathbf{F}^{ - 1}}\omega (p_\tau)\mathbf{F}\frac{{\partial H_\text{qr}(\{\theta(t + \epsilon )\})}}{{\partial {\theta_{i}}}}
   \end{aligned}
\label{eq:eq29}
\end{equation}
In our model, we choose $\omega(p_\tau)$ as,
\begin{equation}
   \omega(p_\tau)=\frac{\max[\sqrt{{2 - 2\cos (\frac{2 \pi p_\tau}{M}) + C}}]}{\sqrt{{2 - 2\cos (\frac{2 \pi p_\tau}{M}) + C}}}
\label{eq:eq30}
\end{equation}
where the numerator is approximately equal to 2, and $C$ is a small non-zero constant and for each $L$ we choose the optimal value of it such that the autocorrelation time is the shortest and at the same it is still finite to avoid $\omega(p_\tau)$ becomes zero. 

As will be shown in Sec.~\ref{sec:ivb}, the FA scheme could greatly cure the critical slowing down in the bare HM scheme, we have succeeded in doing so by applying the Fourier acceleration along the most important direction that dominates the critical fluctuations -- the imaginary time direction.  The DFT is thus performed along the imaginary time direction~\cite{Batrouni2019}. In practice, this means we need to perform DFT $L\times L$ times for every sweep. 

Before the end of this algorithm section, one more point we would like to emphasize is that, although the HM plus FA scheme looks a bit more complicated than the LM, OR and WC schemes, it actually enjoys a big advantage that this method is actually more versatile in terms of implementation for complicated Hamiltonians. In the Hamiltonian mechanics of Eq.~\eqref{eq:eq29}, we actually don't worry too much about the exact form of the Hamiltonian since the updates only depend on its intrinsic dynamics. \pay{In fact, few recent examples implementing the FA scheme in the electron-phonon type of problems in 2d and 3d Holstein models where the fermions and bosons (phonons) are strongly coupled, even with long-range interactions~\cite{Batrouni2019,Bradley2021}, have been successfully shown to reveal the interesting phenomena such as charge-density-wave transition and superconductivity.}    


This completes our discussion of the five different Monte Carlo update algorithms to solve the quantum rotor model and now we are ready to demonstrate our simulation results and check the different performance among the update schemes.

\section{Results}
\label{sec:iv}
In this section, we first use the Monte Carlo method to pin down the precise position of the $(2+1)$d O(2) QCP via finite size scaling analysis, then compare the performance of different update schemes at the QCP.

\begin{figure}[htp!]
	\includegraphics[width=\columnwidth]{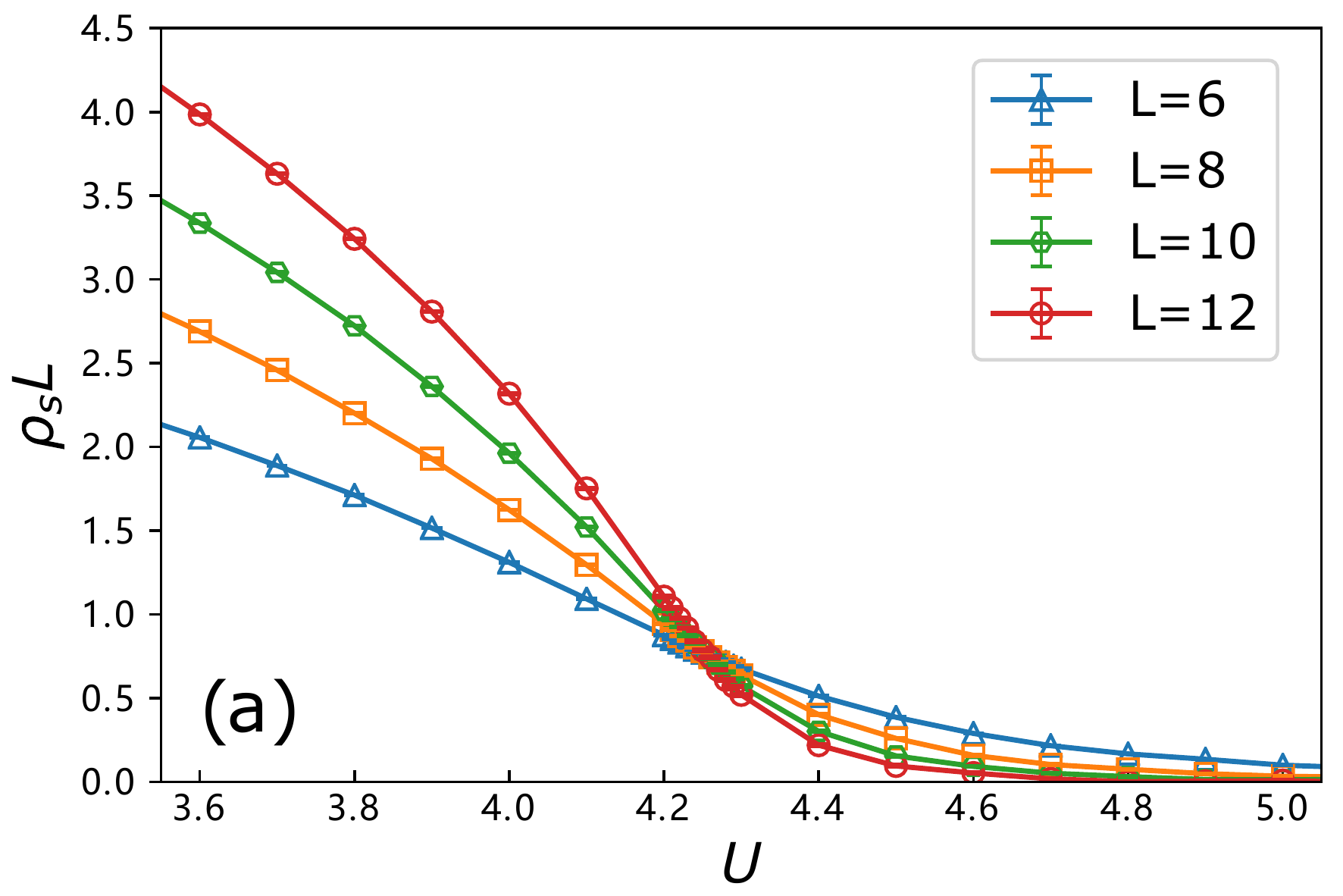}
	\includegraphics[width=\columnwidth]{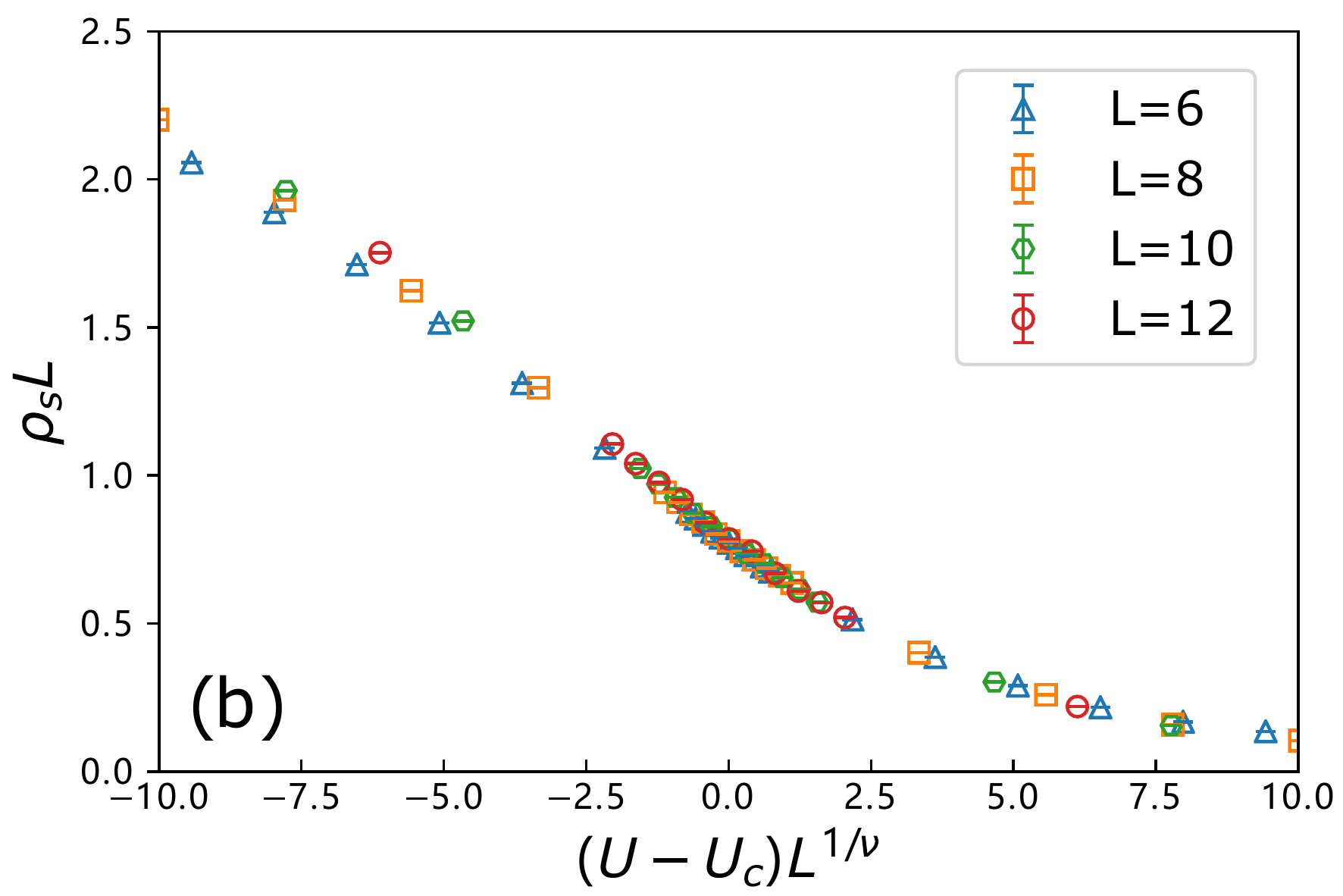}
	\caption{(a) Finite size scaling of the superfluid density according to Eq.~\eqref{eq:eq32}, the crossing point determines the $(2+1)$d XY quantum critical point at $g_c=(\frac{U}{t})_c=4.25(2)$. (b) Data collapse of the above data with $g_c$, $z=1$ and $\nu=0.6723$.}
	\label{fig:fig2}
\end{figure}

\subsection{$(2+1)$d XY quantum phase transition}
\label{sec:iva}
We first identify the position of the quantum critical point of the quantum rotor model in Eqs.~\eqref{eq:eq2}, \eqref{eq:eq12} and \eqref{eq:eq15}. It can be determined by the finite size scaling analysis of the spin stiffness (in the spin language) or superfluid density (in the boson language), as,
\begin{equation}
\rho_s = \frac{1}{2NL_\tau}\sum_{\alpha=\hat{x},\hat{y}}\big(\langle H_{\alpha} \rangle - \langle I^{2}_\alpha\rangle \big),
\label{eq:eq31}
\end{equation}
where $H_{\alpha} = t\Delta \tau \sum_{i,l}\cos(\theta_i(l)-\theta_{i+\alpha}(l))$ is the kinetic energy of the nearest neighbor bond of both spatial directions, and $I_{\alpha}=t\Delta \tau\sum_{i,l}\sin(\theta_i(l)-\theta_{i+\alpha}(l))$ is the derivative of $H_\alpha$. According to the $(2+1)$d XY transition, $\rho_s$ will follow the finite-size scaling function,
\begin{equation}
\rho_s = L^{-z}f((g-g_c)L^{\frac{1}{\nu}}),
\label{eq:eq32}
\end{equation}
with $z=1$ the dynamic exponent of the $(2+1)$d XY universality and $g=\frac{U}{t}$ is the dimensionless control parameter of the transition and $\nu=0.67$ is the correlation length exponent of the $(2+1)$d XY transition~\cite{Hasenbusch1999,Campostrini2001,Campostrini2006,Meng2008,Hasenbusch2019}.

Our results of the superfluid density are shown in Fig.~\ref{fig:fig2}. In Fig.~\ref{fig:fig2} (a) we plot $\rho_s L$ vs $\frac{U}{t}$, and the simulations are performed with $\beta=L$ to make sure we are approaching the quantum critical point. The crossing among different system sizes clearly demonstrates the position of the transition at $g_c=(\frac{U}{t})_c=4.25(2)$ with $\Delta \tau=0.1$ which is well consistent with that in previous literatures~\cite{Wallin1994}. Fig.~\ref{fig:fig2}(b) depicts the data collapse by rescaling the $x$-axis as $(g-g_c)L^{\frac{1}{\nu}}$ with 3d XY exponents, the collapse is in very good quality. 
 
\begin{figure}[htp!]
 \includegraphics[width=\columnwidth]{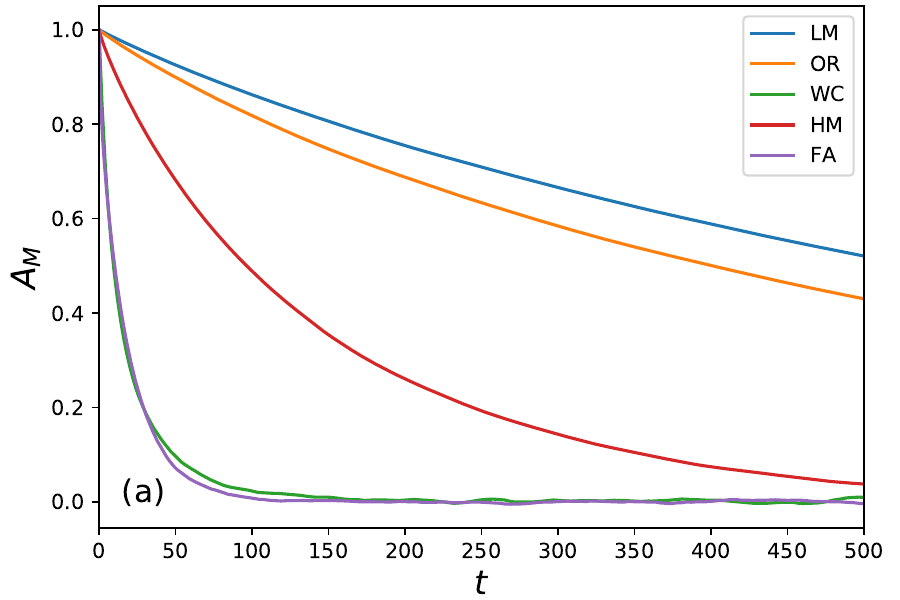}
 \includegraphics[width=\columnwidth]{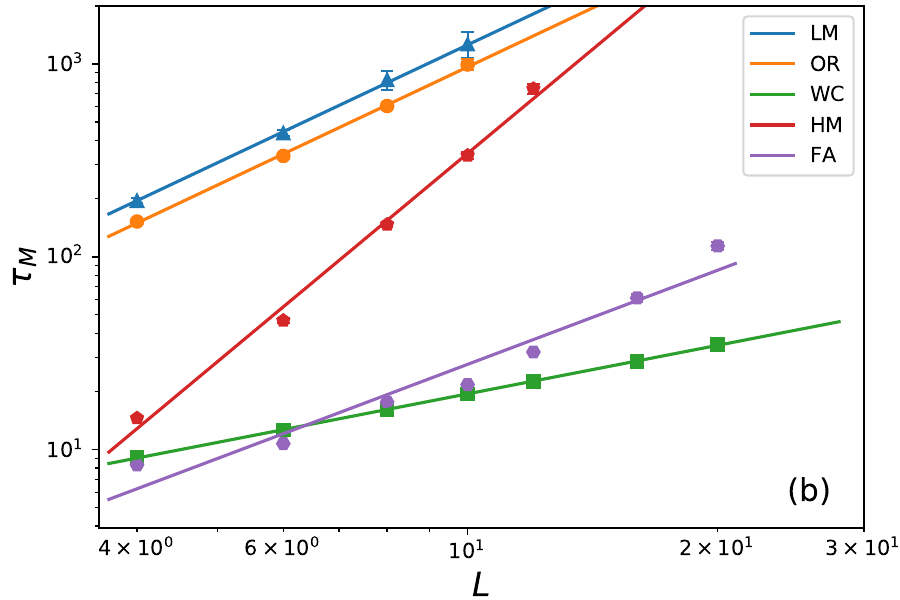}
 \caption{(a) Autocorrelation function of static uniform magnetization $A_M(t)$ in Eq.~\eqref{eq:eq34} at the quantum rotor QCP for LM, OR, WC, HM and FA update schemes with system size $L=8$ and the Monte Carlo time $t$ are averaged over one million consecutive measurements from a single Markov chain. One can see that for LM, OR and HM, the autocorrelation time is very long, more than 500 Monte Carlo sweeps. While for WC and FA, the autocorrelation time decreases a lot. (b) log-log plot of the autocorrelation time $\tau_M$ vs $L$ for magnetization at the quantum rotor QCP for LM, OR, WC, HM and FA update schemes. We fit the data with power-law as in Eq.~\eqref{eq:eq35}, to obtain the corresponding Monte Carlo dynamic exponent $z$. As for the FA scheme, we choose $C$ in Eq.~\eqref{eq:eq30} to be 0.1 for $L=4,6$, 0.01 for $L=8,10$, and 0.005 for $L=12,16,20$, to obtain the optimal value of $\tau_M$.}
 	\label{fig:fig3}
 \end{figure}

\begin{figure}[htp!]
   \includegraphics[width=\columnwidth]{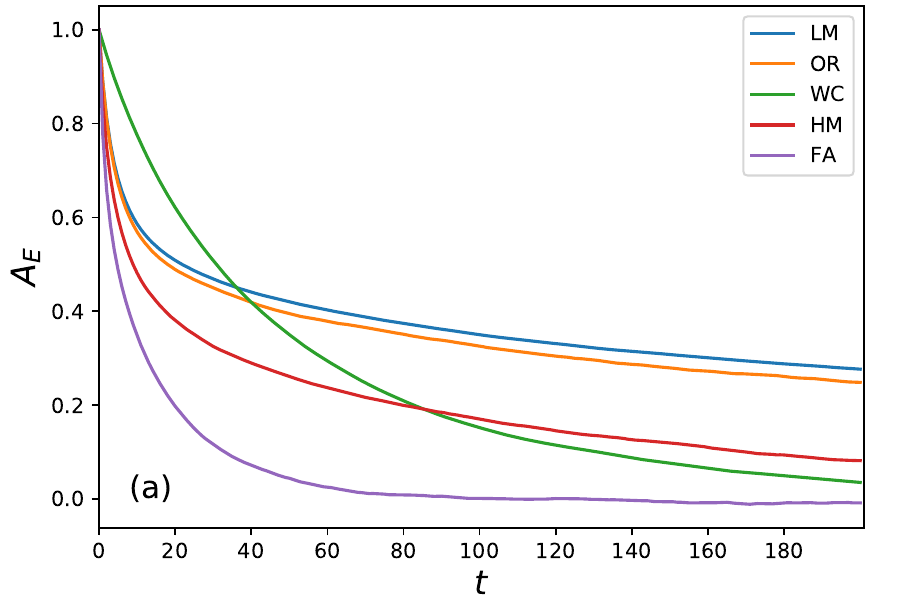}
   \includegraphics[width=\columnwidth]{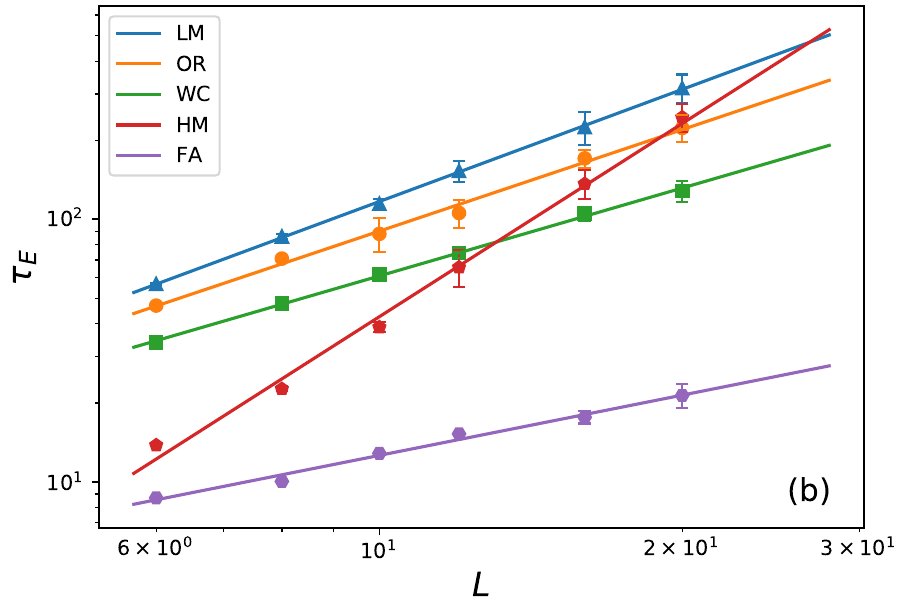}
	\caption{(a) Autocorrelation function of energy $A_{E}(t)$ at the quantum rotor QCP for LM, OR, WC, HM and FA update schemes with system size $L=8$ and the Monte Carlo time $t$ are averaged over one million consecutive measurements from a single Markov chain. One can see that for LM, OR and HM, the autocorrelation time is very long, more than 200 Monte Carlo sweeps. While for WC and FA, the autocorrelation time decreases a lot. (b) log-log plot of the autocorrelation time $\tau_E$ for energy at the quantum rotor QCP for LM, OR, WC, HM and FA update schemes. Again we fit the data with power-law as in Eq.~\eqref{eq:eq35}, to obtain the corresponding Monte Carlo dynamic exponent $z$. As for the FA scheme, we choose $C$ in Eq.~\eqref{eq:eq30} to be 0.1 for $L=6,8,10,12$, and 0.005 for $L=16,20$, to obtain the optimal value of $\tau_E$.} 
	\label{fig:fig4}
\end{figure}

\subsection{Autocorrelation time analysis}
\label{sec:ivb}
With the $g_c$ determined, we can now explore the performance of various update schemes in the vicinity of the critical point. We first analyze the autocorrelation time of the static uniform magnetization $M$ at the QCP,
\begin{equation}
M= \frac{|\sum_{\mathbf{r}} \vec{\theta}_{\mathbf{r}} |} {V},
\label{eq:eq33}
\end{equation}
$\theta_\mathbf{r}$ is an unit vector here with the phase between $0$ to $2\pi$. The summation is performed over the space-time volume, therefore the $M$ is the static uniform magnetization of the O(2) order parameter (see Fig.~\ref{fig:fig1} for schematics). We measured the magnetization right at the QCP $g=g_c$, and obtain its autocorrelation time $\tau$ from fitting the exponential decay of the autocorrelation function of $A_M(t)$ as,
\begin{equation}
A_M(t) = \frac{\langle M(t)M(0) \rangle - \langle M(0)\rangle^2}{\langle M^2 \rangle - \langle M \rangle^2} \sim e^{-\frac{t}{\tau}},
\label{eq:eq34}
\end{equation} 
where $t$ is the time in the unit of one Monte Carlo sweep, although the definition of one sweep can be slightly different among the update schemes, basically it corresponds to one complete update of the space-time lattice. The results are shown in Fig.~\ref{fig:fig3}. To obtain these smooth autocorrelation functions, we use $10^6$ Monte Carlo measurements in a single Markov chain to calculate the $A_{M}(t)$. From Fig.~\ref{fig:fig3} (a), the exponential decay of the autocorrelation functions for $L=8$, $\beta=L$ at the QCP are clearly visible. Then one read the autocorrelation time from such results for different $L$. These results show that LM, OR and HM schemes are all suffering from the critical slowing down and the autocorrelation time increases drastically with the system size $L$. On the other hand, WC and FA have very small autocorrelation time and are for sure the suitable methods to apply here. To quantify the difference in the autocorrelation time, we take the log-log plot in Fig.~\ref{fig:fig3} (b), and expect a power-law relation of the form, 
\begin{equation}
\tau \sim L^{z}
\label{eq:eq35}
\end{equation}
where $z$ the dynamical exponent of the Monte Carlo update scheme, and for the 2d Ising model at its critical point, it is known that the $z=2.2$ for the local update and $z=0.2$ for the Swendsen-Wang cluster update~\cite{SwendsenWang1987}. From Fig.~\ref{fig:fig3} (b) one can read the $z=2.05(16)$ for LM, $z=2.05(8)$ for OR, and $z=3.60(5)$ for HM, and as for the other two update schemes, shorter autocorrelation times are observed, for example, $z=0.84(2)$ for WC, and $z=1.62(30)$ for FA. So these results reveal that at the QCP of $(2+1)$d quantum rotor model, the critical slowing down is mostly suppressed in WC and FA schemes.
 
The same analysis can be performed for the autocorrelation time of energy at the QCP, and the results are shown in Fig.~\ref{fig:fig4}, one can read the $z=1.40(11)$ for LM, $z=1.28(8)$ for OR, and $z=2.43(12)$ for HM. As for the other two update schemes, shorter autocorrelation times are observed, for example, $z=1.11(6)$ for WC, and most importantly, $z=0.76(8)$ for FA with the smallest autocorrelation time of all sizes. It illustrates that the autocorrelation time for energy is different from that for the magnetization -- it is actually normal since different physical observables can have different autocorrelation time -- these results nevertheless reveal the consistent picture that at the QCP of $(2+1)$d quantum rotor model, the critical slowing down is mostly suppressed in WC and FA schemes. 


\subsection{CPU time analysis}
\label{sec:ivc}
Besides the autocorrelation time and its scaling with system size at the QCP, we also test the effective calculation time of each Monte Carlo scheme. 

\begin{figure}[htp!]
	\includegraphics[width=\columnwidth]{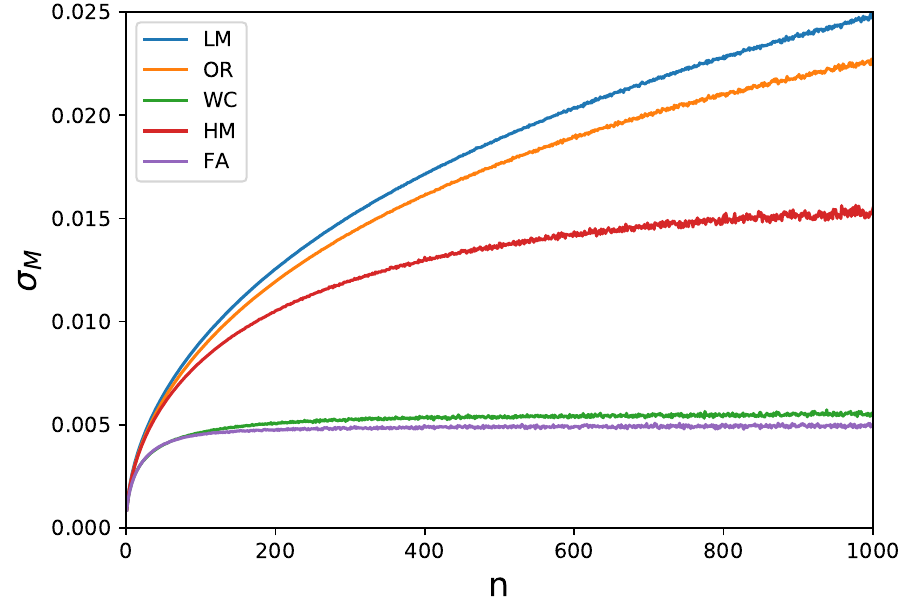}
	\caption{The rebinning of the static uniform magnetization of different algorithms as a function of bin size $n$. We choose the $L=8$ and $\beta=8$ system at the QCP, and rebin the consecutive measurements to obtain the intrinsic errorbars $\sigma_M$ of different update schemes. It is clear for WC and FA, the plateaux of errorbar are already reached when $n<200$, and for the other update schemes, it will take much longer CPU time for the intrinsic errorbar plateaux to be reached, if ever reached.}
	\label{fig:fig5}
\end{figure}

For example, for the magnetization at $L=8$, $\beta=L$ at the QCP, we compute the real CPU time and the obtained errorbars among different schemes. 
Here, we use method of rebinning to estimate the errorbar of data for each update scheme and the time it takes to reach that. As shown in Fig.~\ref{fig:fig5}, with fixed sample number of magnetization, we group the data of every $n$ consecutive measurements into one bin, then calculate errorbar of sample mean among these bins. As the bin size $n$ becomes large, the correlation of sample mean among the bins becomes small, and an plateau of the errorbar will be reached once the data among different bins are indeed statistically independent. Among different schemes and it is clear that for WC and FA scheme, not only the plateau in errorbar are reach at the earliest, around $n\sim 100$ of the bin size, but also the intrinsic errorbars obtained in this way are actually the smallest among the five update schemes. These results implies the WC and FA schemes will be able to acquire the best quality data with the smallest intrinsic errorbars of the magnetization at the QCP with the shortest CPU time. All the other schemes, LM, OR and HM, will take much longer time for the intrinsic plateaux of their errorbars to be reached and hence will take longer in CPU time. 

\begin{figure}[htp!]
	\includegraphics[width=\columnwidth]{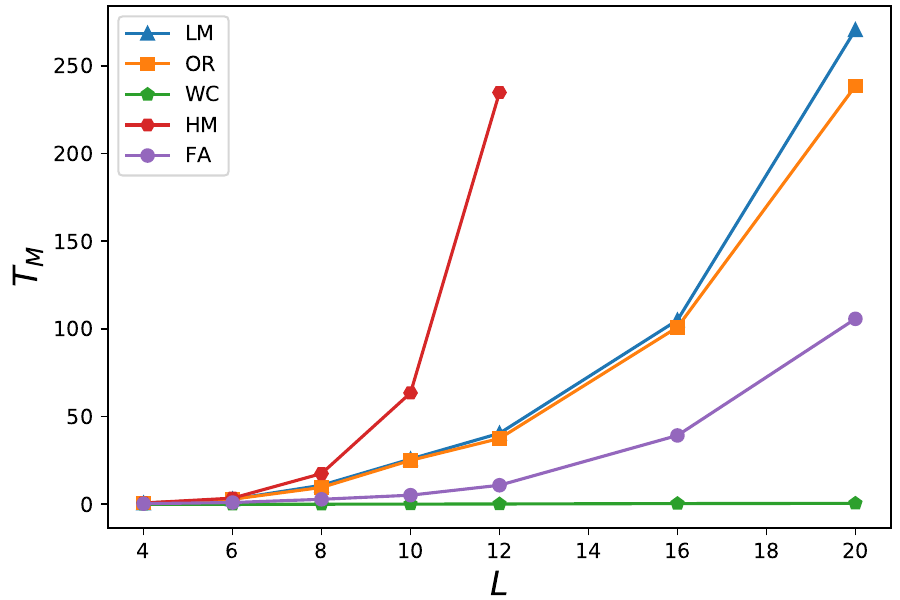}
	\caption{The CPU hour $T_M$ spent for each Monte Carlo scheme to obtain the error of 0.001 of static uniform magnetization $M$, as a function of system sizes $L$. As expected, although the HM scheme has shorter autocorrelation time compared with LM scheme, it actually spends more CPU time on obtaining the shorter autocorrelation time mainly due to the leapfrog processes therein. Whereas with FA scheme, both the autocorrelation time and the CPU hour spent are much less compared with LM, OR, HM schemes in reaching the same errorbar. The less CPU hours spent is for the WC scheme, owing to both the small autocorrelation time and less number of operations in the algorithm, thus it will be the best to calculate the bigger sizes in the quantum rotor model.}
	\label{fig:fig6}
\end{figure}

The actual CPU hour spent on achieving the errorbar of order $0.001$ for the static uniform magnetization are shown in Fig.~\ref{fig:fig6}. As explained above, it is indeed the case that the HM, LM, OR schemes need longer physical time to achieve the required level of statistical error and among these three, the FA scheme is the slowest mainly due to the leapfrog processes therein, although it shows smaller $\tau_M$ and $\tau_E$ compared with LM and OR schemes. The FA and WC schemes, on the contrary, need very small CPU hours to achieve the required error and the WC is the best in this regard. But it should be mentioned that although the FA scheme spent longer physical time than WC mainly due to its computational complextiy. It introduces more adjustable coefficients such as $N_{\text{HM}}$, $\epsilon$, and $\omega(p_\tau)$, and it offers the opportunity of efficient global update even if the Hamiltonian of the problem at hand are more complicated, such as the Fermion-boson coupled systems aforementioned, where usually long-range interactions are present and one can no longer construct WC type of update.


\section{Discussion}
\label{sec:v}
In this paper, we systematically test the performance of several Monte Carlo update schemes for the $(2+1)$d O(2) phase transition of quantum rotor model. Our results reveal that comparing the local Metropolis (LM), LM plus over-relaxation (OR), Wolff-cluster (WC), hybrid Monte Carlo (HM), hybrid Monte Carlo with Fourier acceleration (FA) schemes, it is clear that, at the quantum critical point, the WC and FA schemes acquire the shortest autocorrelation time and cost the least amount of CPU time in achieving the smallest level of relative error. 

As we have repeatedly discussed throughout the paper, although the $(2+1)$d quantum rotor models have been satisfactorily solved with various analytic (such as high temperature expansion~\cite{Campostrini2001,Campostrini2006}, conformal bootstrap~\cite{Chester2019}) and Monte Carlo simulation schemes, it is now becoming more clear to the community that the extension of quantum rotor model to more realistic and yet challenging models, such as quantum rotor models -- playing the role of ferromagnetic/antiferromagnetic critical bosons~\cite{XYXu2017,ZHLiu2019}, Z$_2$ topological order~\cite{Senthil2002} and U(1) gauge field in QED$_3$~\cite{XYXu2019,WangWei2019} -- Yukawa-coupled to various Fermi surfaces will provide the key information upon the important and yet unsolved physical phenomena ranging from non-Fermi-liquid, reconstruction of Fermi surface beyond the Luttinger theorem~\cite{Paramekanti2004,ChuangChen2019,Gazit2019,ChuangChen2021} and whether the monopole operator is relevant or irrelevant at QED$_3$ with matter field~\cite{Dupuis2019,WangWei2019}. Furthermore, the bosonic fluctuation may even be applied in the momentum space, e.g., the Twisted Bilayer Graphene~\cite{Xu2021}. And progresses in the problem of Fermi surface Yukawa-coupled to the quantum rotor model~\cite{weilunjiang2021pseudogap,yuzhiliu2021dynamical} have been made by some of us, in which, we combine three of five update schemes discussed in the present work, and show that they together give us a general pattern in effectively sampling the O(2) bosons in the coupled systems. Novel physical phenomena including the non-Fermi-liquid quantum critical metal, the deconfined critical point~\cite{Anders2020}, and the pseudogap and superconducting phases which are generated from the quantum critical bosonic fluctuations, are discovered from QMC simulations.

Moreover, although the fundamentalness of these questions go way beyond the simple $(2+1)$d Wilson-Fisher O(2) fixed point, but the successful solution of these questions heavily relies on the design of more efficient Monte Carlo update schemes on the quantum rotor or O(2) degree of freedoms in the $(2+1)$d configuration space of the aforementioned problem. In the presence of the fermion determinant, one can perform the traditional determinantal quantum Monte Carlo method with local update of the O(2) rotors for small the medium system sizes, then with the available self-learning and neural network schemes~\cite{JWLiu2017,XYXuSelf2017,HYLu2021}, an effective model with non-local interactions among the rotors can be obtained which serves as the low-energy description of the fermion-boson coupled systems. Then the methods tested in this paper, in particular the hybrid Monte Carlo with Fourier acceleration scheme, can be readily employed the perform global update for such effective model, which will certainly reduce the autocorrelation time compared with the simulation of the original model and consequently reduce the actual CPU hours spent in achieving the same level of numerical accuracy of the physical observables. \pay{Recent progress in the Holstein-type problems, where fermions and phonons (bosons) are strongly coupled in 2d and 3d lattices with long-range interactions with FA scheme, has been shown to be successful in revealing various charge-density-wave and superconductivity transitions~\cite{Batrouni2019,Bradley2021}.}

\section*{Acknowledgement}
We thank Ying-Jer Kao for insightful discussion and introduction to us the over-relaxation update scheme in Ref.~\cite{Lan2012}. WLJ, GPP, YZL and ZYM acknowledge the supports from the Strategic Priority Research Program of the Chinese Academy of Sciences (Grant No. XDB33000000) and the RGC of Hong Kong SAR of China (Grant Nos. 17303019, 17301420 and AoE/P-701/20). We thank the Center for Quantum Simulation Sciences in the Institute of Physics, Chinese Academy of Sciences, the Computational Initiative at the Faculty of Science at the University of Hong Kong and the National Supercomputer Center in Tianjin and the National Supercomputer Center in Guangzhou for their technical support and generous allocation of CPU time.

\begin{appendix}
\section{link-current representation scheme}
   \label{app:appA}
From partition function in Eq.~\eqref{eq:eq12}, one can also integrate $\theta$ to arrive at the following link-current model. Its partition function is written as, 
   \begin{equation}
   Z = \sum\limits_{\{ J\} }\exp [  - \frac{1}{2}\sum\limits_{\mathbf{r}} {\sum\limits_{\mathbf{\mu}  = x,y,\tau } {{K_{ \langle \mathbf{r},\mathbf{r} + \mathbf{\mu}  \rangle }}{{({J_{\mathbf{r},\mathbf{\mu}}})}^2}} } ] 
   \label{eq:eqA1}
   \end{equation}
where $\mathbf{r}$ represents the original site in the $(2+1)$d space-time configuration, $\mathbf{\mu}$ represents the $\pm x,\pm y,\pm \tau$ bond directions originating from site $\mathbf{r}$. $J_{\mathbf{r},\mathbf{\mu}}$ lies on the bond $\langle \mathbf{r},\mathbf{r}+\mu \rangle$. For spatial bonds $K_{\langle \mathbf{r},\mathbf{r'}\rangle}=K_x=t \Delta \tau$ and for temporal bonds $K_{\langle \mathbf{r},\mathbf{r'}\rangle} = K_\tau = \frac{1}{U \Delta \tau}$. Due to the translation symmetry, $J_{\mathbf{r},\mathbf{\mu}} = -J_{\mathbf{r}+\mathbf{\mu},-\mathbf{\mu}}$, then the current on bonds must obey continuity equation,
\begin{equation}
\sum\limits_\mathbf{\mu}  {{J_{\mathbf{r},\mathbf{\mu} }}}  = 0.
   \label{eq:eqA2}
\end{equation}
One can update model Eq.~\eqref{eq:eqA1} in a loop update regime~\cite{Wallin1994,Alet2003}. By choosing one loop with probability like worm algorithm in term of the weight, $A_{\mathbf{r},\mathbf{\mu}}$, written as,
   \begin{equation}
      A_{\mathbf{r},\mathbf{\mu}} = \min [1,\exp ( - \Delta E_{\mathbf{r},\mathbf{\mu}})],
      \label{eq:eqA3}
   \end{equation}
where $\Delta E_{\mathbf{r}} = E'_{\mathbf{r}} - E_{\mathbf{r}}$ and $E'_{\mathbf{r}}$ and $E_{\mathbf{r}}$ represent the classical energy after and before the update on bond $\langle \mathbf{r},\mathbf{r}+\mu \rangle$ described by Eq.~\eqref{eq:eqA1}. Thus the probability can be derived by normalize the weight, $p_{\mathbf{r},\mathbf{\mu}} = \frac{A_{\mathbf{r},\mathbf{\mu}}}{N_{\mathbf{r}}}$, where $N_{\mathbf{r}}=\sum\limits_\mathbf{\mu} {A_{\mathbf{r},\mathbf{\mu}}}$. Usually we add the same integer to all bonds on the loop to satisfy the continuity equation Eq.~\eqref{eq:eqA2}. 

Moreover, transferring the lattice to its dual one can avoid the constraint, as the Monte Carlo update with constraint is usually difficult to implement. One can easily update one site instead of a loop, by writing,
   \begin{equation}
      \begin{aligned}
            {J_{\mathbf{r},x}} &= {M_{\mathbf{r} + x}} + {M_{\mathbf{r} + x + y}} + {M_{\mathbf{r} + x + z}} + {M_{\mathbf{r} + x + y + z}}\\
            {J_{\mathbf{r},y}} &= {M_{\mathbf{r} + y}} + {M_{\mathbf{r} + y + x}} + {M_{\mathbf{r} + y + z}} + {M_{\mathbf{r} + y + x + z}}\\
            {J_{\mathbf{r},\tau }} &=  - 2({M_{\mathbf{r} + z}} + {M_{\mathbf{r} + z + x}} + {M_{\mathbf{r} + z + y}} + {M_{\mathbf{r} + z + x + y}})\\
            {J_{\mathbf{r},-x}} &= {M_\mathbf{r}} + {M_{\mathbf{r} + y}} + {M_{\mathbf{r} + z}} + {M_{\mathbf{r} + y + z}}\\
            {J_{\mathbf{r},-y}} &= {M_\mathbf{r}} + {M_{\mathbf{r} + x}} + {M_{\mathbf{r} + z}} + {M_{\mathbf{r} + x + z}}\\
            {J_{\mathbf{r},-\tau }} &=  - 2({M_\mathbf{r}} + {M_{\mathbf{r} + x}} + {M_{\mathbf{r} + y}} + {M_{\mathbf{r} + x + y}})
      \end{aligned}
      \label{eq:eqA4}
   \end{equation}

\clearpage
\begin{figure}[htbp!]
   \includegraphics[width=\columnwidth]{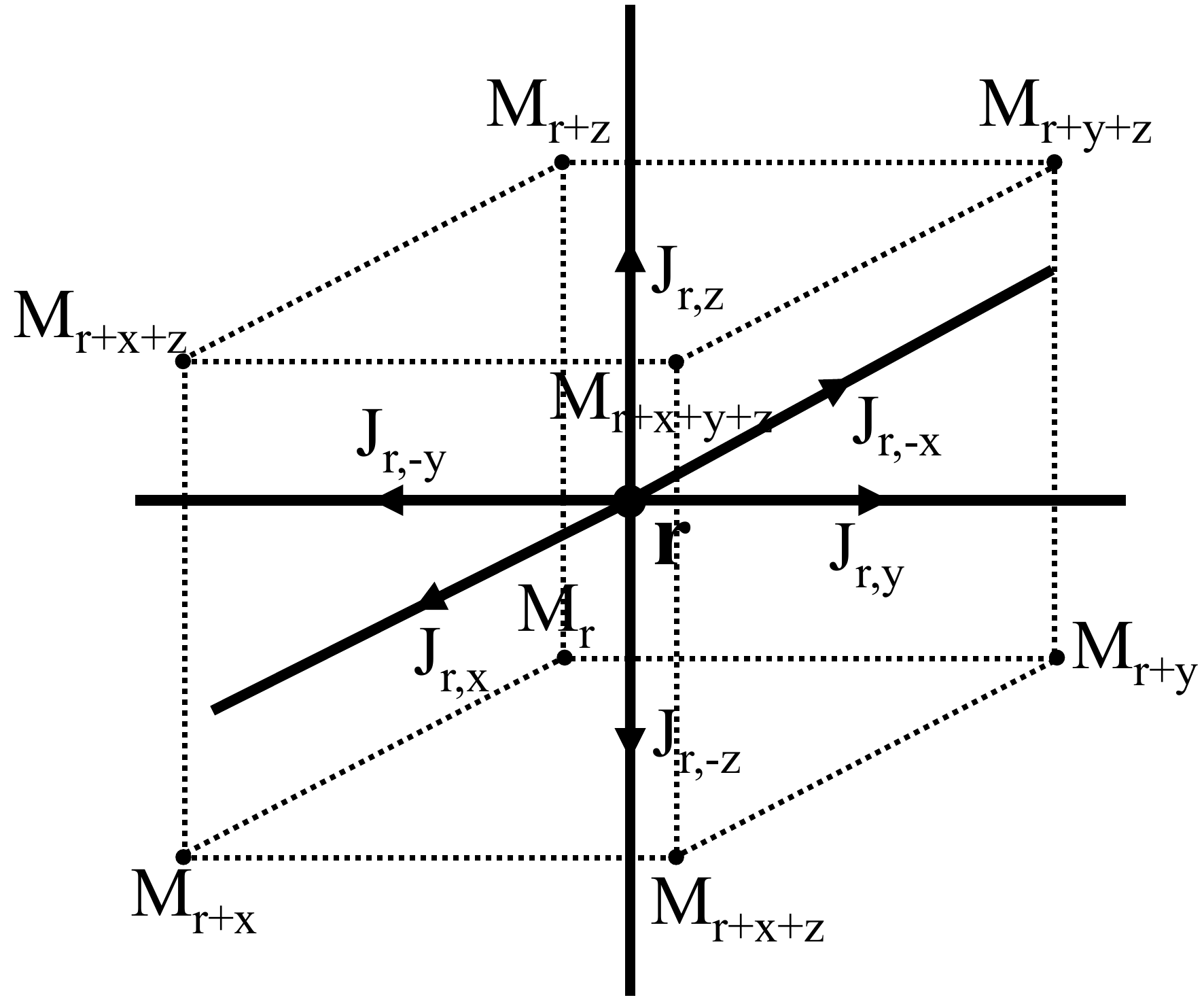}
   \caption{The map between the original lattice to dual lattice, according to Eq.~\eqref{eq:eqA4}.}
   \label{fig:fig7}
\end{figure}

As denoted in Fig.~\ref{fig:fig7}, that ${J_{\mathbf{r},\mu}}$ lives on the original bond of square lattice on the site $\mathbf{r}$ with direction $\mu$ and $M_{\mathbf{r}}$ represents the site of dual lattice. The continuity equation is naturally satisfied. Now the partition function can be written as,
\begin{equation}
      \begin{aligned}
      Z = &\sum\limits_{\square_{x,y}} {\exp [ - \frac{1}{2}} {K_x}{({M_\mathbf{r}} + {M_{\mathbf{r} + \mu }} + {M_{\mathbf{r} + z}} + {M_{\mathbf{r} + \mu  + z}})^2}] \\ 
      + &\sum\limits_{\square_{z}} {\exp [ - 2{K_\tau }{{({M_\mathbf{r}} + {M_{\mathbf{r} + x}} + {M_{\mathbf{r} + y}} + {M_{\mathbf{r} + x + y}})}^2}]}
      \end{aligned}
   \label{eq:eqA5}
\end{equation}

   Here, $\mu$ represnets $x$ and $y$ direction. The summation in the first line of Eq.~\eqref{eq:eqA5} runs over all the square whose normal vector is in the direction of $x$ and $y$.
   Similarly, The summation in the second line of Eq.~\eqref{eq:eqA5} runs over all the square whose normal vector is in the direction of $z$. Local update can be easily used in this situation of one site by simply calculating the energy difference of the square it locates.

\vspace*{190pt}

\section{WC update scheme}
\label{app:appB}
\begin{figure}[htbp!]
   \includegraphics[width=\columnwidth]{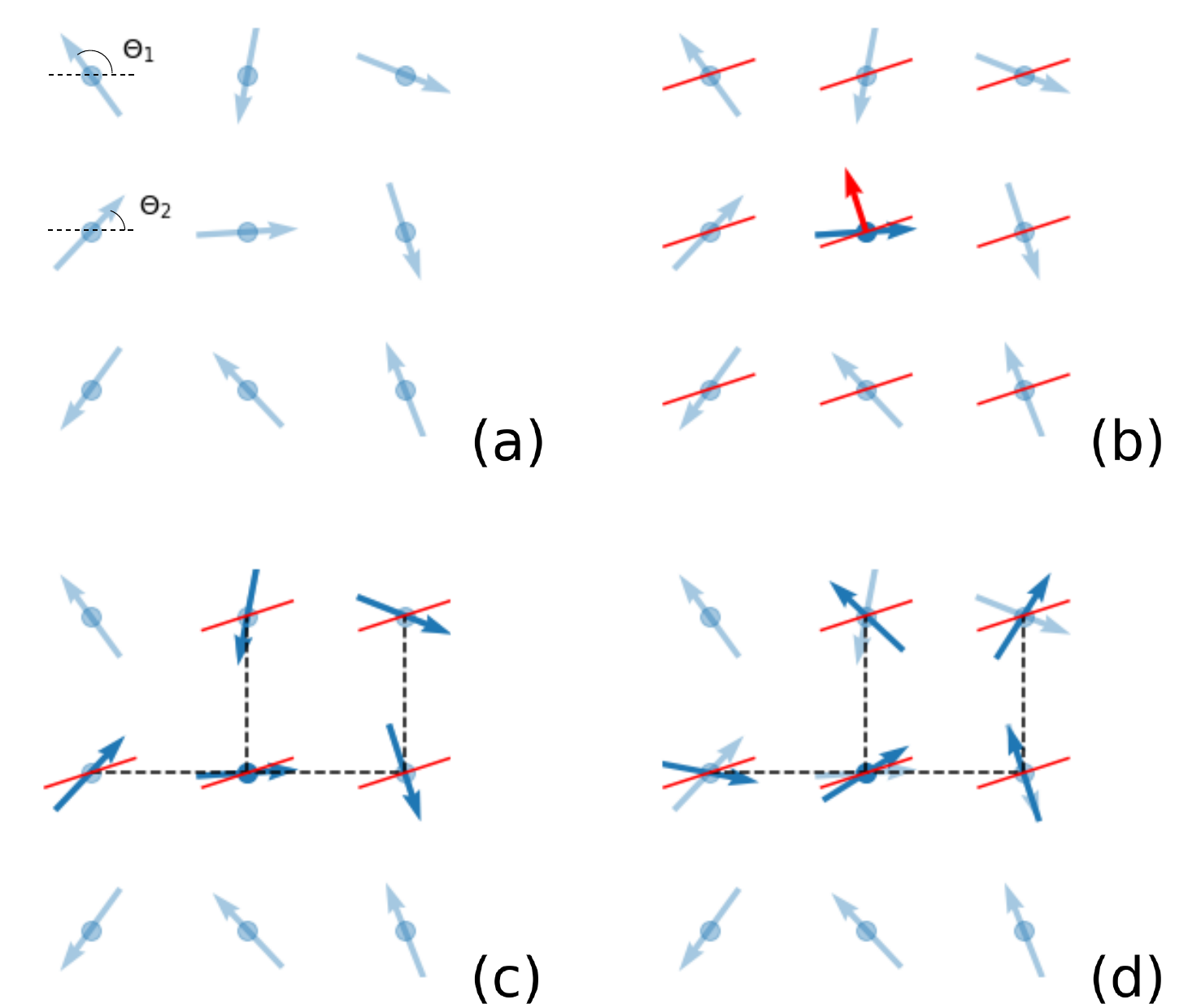}
   \centering
   \caption{Schematic map of WC update operation. (a) The light blue sites and arrows represent the lattice and the orientations, with the angle of $\theta$. (b) The red short lines for each sites represents the common mirror direction, whose normal vector is denoted by the red arrow. The deep blue arrow in the middle is the vector on the random-choosing site, served as the start point of the cluster. (c) The deep blue arrows linked by the dash lines denote the vector in the cluster, constructed by probability. (d) The deep blue arrows here are the reoriented vectors, compared with the light blue arrows before update.}
      \label{fig:fig8}
\end{figure}

Here we show the detailed update process by means of part of sites in the whole system. Four subgraph in Fig.~\ref{fig:fig8} show the vectors before update, choosing the random site and mirror denoted by $\hat r$, construct the cluster with probability, reorient the vectors, respectively.

\end{appendix}

\newpage
\bibliographystyle{iopart-num.bst}
\bibliography{Note}

\providecommand{\newblock}{}
\begin{thebibliography}{10}
\expandafter\ifx\csname url\endcsname\relax
  \def\url#1{{\tt #1}}\fi
\expandafter\ifx\csname urlprefix\endcsname\relax\def\urlprefix{URL }\fi
\providecommand{\eprint}[2][]{\url{#2}}

\bibitem{Jose1977}
Jos\'e J~V, Kadanoff L~P, Kirkpatrick S and Nelson D~R 1977 {\em Phys. Rev.
  B\/} {\bf 16} 1217

\bibitem{Wallin1994}
Wallin M, So/rensen E~S, Girvin S~M and Young A~P 1994 {\em Phys. Rev. B\/}
  {\bf 49} 12115

\bibitem{Hasenbusch1999}
Hasenbusch M and T\''or\''ok T 1999 {\em Journal of Physics A: Mathematical and
  General\/} {\bf 32} 6361

\bibitem{Hasenbusch2019}
Hasenbusch M 2019 {\em Phys. Rev. B\/} {\bf 100} 224517

\bibitem{Lan2012}
{Lan} T~Y, {Hsieh} Y~D and {Kao} Y~J 2012
  arXiv:1211.0780[cond--mat.stat--mech]

\bibitem{WanwanXu2019}
Xu W, Sun Y, Lv J~P and Deng Y 2019 {\em Phys. Rev. B\/} {\bf 100} 064525

\bibitem{Campostrini2001}
Campostrini M, Hasenbusch M, Pelissetto A, Rossi P and Vicari E 2001 {\em Phys.
  Rev. B\/} {\bf 63} 214503

\bibitem{Campostrini2006}
Campostrini M, Hasenbusch M, Pelissetto A and Vicari E 2006 {\em Phys. Rev.
  B\/} {\bf 74} 144506

\bibitem{Chester2019}
Chester S~M, Landry W, Liu J, Poland D, Simmons-Duffin D, Su N and Vichi A 2020
  {\em Journal of High Energy Physics\/} {\bf 2020} 1

\bibitem{Fisher1988}
Fisher M~P~A and Grinstein G 1988 {\em Phys. Rev. Lett.\/} {\bf 60} 208

\bibitem{Cha1991}
Cha M~C, Fisher M~P~A, Girvin S~M, Wallin M and Young A~P 1991 {\em Phys. Rev.
  B\/} {\bf 44} 6883

\bibitem{Fisher1989}
Fisher M~P~A, Weichman P~B, Grinstein G and Fisher D~S 1989 {\em Phys. Rev.
  B\/} {\bf 40} 546

\bibitem{Greiner2002}
Greiner M, Mandel O, Esslinger T, H\''ansch T~W and Bloch I 2002 {\em Nature\/}
  {\bf 415} 39

\bibitem{Meng2008}
Meng Z~Y and Wessel S 2008 {\em Phys. Rev. B\/} {\bf 78} 224416

\bibitem{HanLi2019}
Li H, Liao Y~D, Chen B~B, Zeng X~T, Sheng X~L, Qi Y, Meng Z~Y and Li W 2020
  {\em Nature communications\/} {\bf 11} 1

\bibitem{ZeHu2020}
Hu Z, Ma Z, Liao Y~D, Li H, Ma C, Cui Y, Shangguan Y, Huang Z, Qi Y, Li W, Meng
  Z~Y, Wen J and Yu W 2020 {\em Nature Communications\/} {\bf 11} 5631

\bibitem{YDLiao2021}
Da~Liao Y, Li H, Yan Z, Wei H~T, Li W, Qi Y and Meng Z~Y 2021 {\em Phys. Rev.
  B\/} {\bf 103} 104416

\bibitem{SwendsenWang1987}
Swendsen R~H and Wang J~S 1987 {\em Phys. Rev. Lett.\/} {\bf 58} 86

\bibitem{Wolff1989}
Wolff U 1989 {\em Phys. Rev. Lett.\/} {\bf 62} 361

\bibitem{Adler1981}
Adler S~L 1981 {\em Phys. Rev. D\/} {\bf 23} 2901

\bibitem{Alet2003}
Alet F and S\o{}rensen E~S 2003 {\em Phys. Rev. E\/} {\bf 68} 026702

\bibitem{XYXu2017}
Xu X~Y, Sun K, Schattner Y, Berg E and Meng Z~Y 2017 {\em Phys. Rev. X\/} {\bf
  7} 031058

\bibitem{ZHLiu2019}
Liu Z~H, Pan G, Xu X~Y, Sun K and Meng Z~Y 2019 {\em Proc. Natl. Acad. Sci.\/}
  {\bf 116} 16760

\bibitem{XYXu2019}
Xu X~Y, Qi Y, Zhang L, Assaad F~F, Xu C and Meng Z~Y 2019 {\em Phys. Rev. X\/}
  {\bf 9} 021022

\bibitem{WangWei2019}
Wang W, Lu D~C, Xu X~Y, You Y~Z and Meng Z~Y 2019 {\em Phys. Rev. B\/} {\bf
  100} 085123

\bibitem{XYXuReview2019}
Xu X~Y, Liu Z~H, Pan G, Qi Y, Sun K and Meng Z~Y 2019 {\em Journal of Physics:
  Condensed Matter\/} {\bf 31} 463001

\bibitem{JWLiu2017}
Liu J, Qi Y, Meng Z~Y and Fu L 2017 {\em Phys. Rev. B\/} {\bf 95} 041101

\bibitem{XYXuSelf2017}
Xu X~Y, Qi Y, Liu J, Fu L and Meng Z~Y 2017 {\em Phys. Rev. B\/} {\bf 96}
  041119

\bibitem{HYLu2021}
{Lu} H, {Li} C, {Li} W, {Qi} Y and {Meng} Z~Y 2021   arXiv:2106.00712
  [cond--mat.str--el]

\bibitem{Fisher1989_2}
Fisher M~P~A 1989 {\em Phys. Rev. Lett.\/} {\bf 62} 1415

\bibitem{Metropolis1953}
Metropolis N, Rosenbluth A~W, Rosenbluth M~N, Teller A~H and Teller E 1953 {\em
  The Journal of Chemical Physics\/} {\bf 21} 1087

\bibitem{Davies1986}
Davies C, Batrouni G, Katz G, Kronfeld A, Lepage P, Rossi P, Svetitsky B and
  Wilson K 1986 {\em Journal of Statistical Physics\/} {\bf 43} 1073

\bibitem{Ferreira1993}
Ferreira A~L and Toral R 1993 {\em Phys. Rev. E\/} {\bf 47} R3848

\bibitem{Duane1988Hybrid}
Duane S, Kennedy A~D, Pendleton B~J and Roweth D 1988 {\em Physics Letters B\/}
  {\bf 195} 216

\bibitem{Simon2002Testing}
Catterall S and Karamov S 2002 {\em Physics Letters B\/} {\bf 528} 301

\bibitem{Batrouni1985}
Batrouni G~G, Katz G~R, Kronfeld A~S, Lepage G~P, Svetitsky B and Wilson K~G
  1985 {\em Phys. Rev. D\/} {\bf 32} 2736

\bibitem{Hastings1970}
Hastings W~K 1970 {\em Biometrika\/} {\bf 57} 97

\bibitem{Whitmer1984}
Whitmer C 1984 {\em Phys. Rev. D\/} {\bf 29} 306

\bibitem{DUANE1987216}
Duane S, Kennedy A, Pendleton B~J and Roweth D 1987 {\em Physics Letters B\/}
  {\bf 195} 216

\bibitem{Gupta1998}
Gupta R, Kilcup G~W and Sharpe S~R 1988 {\em Phys. Rev. D\/} {\bf 38} 1278

\bibitem{Mehlig1992}
Mehlig B, Heermann D~W and Forrest B~M 1992 {\em Phys. Rev. B\/} {\bf 45} 679

\bibitem{Scalettar1987}
Scalettar R~T, Scalapino D~J, Sugar R~L and Toussaint D 1987 {\em Phys. Rev.
  B\/} {\bf 36} 8632

\bibitem{Beyl2018}
Beyl S, Goth F and Assaad F~F 2018 {\em Phys. Rev. B\/} {\bf 97} 085144

\bibitem{Batrouni2019}
Batrouni G~G and Scalettar R~T 2019 {\em Phys. Rev. B\/} {\bf 99} 035114

\bibitem{Bradley2021}
Bradley O, Batrouni G~G and Scalettar R~T 2021 {\em Phys. Rev. B\/} {\bf 103}
  235104

\bibitem{Senthil2002}
Senthil T and Motrunich O 2002 {\em Phys. Rev. B\/} {\bf 66} 205104

\bibitem{Paramekanti2004}
Paramekanti A and Vishwanath A 2004 {\em Phys. Rev. B\/} {\bf 70} 245118

\bibitem{ChuangChen2019}
Chen C, Xu X~Y, Qi Y and Meng Z~Y 2020 {\em Chinese Physics Letters\/} {\bf 37}
  047103

\bibitem{Gazit2019}
Gazit S, Assaad F~F and Sachdev S 2020 {\em Phys. Rev. X\/} {\bf 10} 041057

\bibitem{ChuangChen2021}
Chen C, Yuan T, Qi Y and Meng Z~Y 2021 {\em Phys. Rev. B\/} {\bf 103} 165131

\bibitem{Dupuis2019}
Dupuis E, Paranjape M~B and Witczak-Krempa W 2019 {\em Phys. Rev. B\/} {\bf
  100} 094443

\bibitem{Xu2021}
Zhang X, Pan G, Zhang Y, Kang J and Meng Z~Y 2021 {\em Chinese Physics
  Letters\/} {\bf 38} 077305

\bibitem{weilunjiang2021pseudogap}
Jiang W, Liu Y, Klein A, Wang Y, Sun K, Chubukov A~V and Meng Z~Y 2021
  arXiv:2105.03639[cond--mat.str--el]

\bibitem{yuzhiliu2021dynamical}
Liu Y, Jiang W, Klein A, Wang Y, Sun K, Chubukov A~V and Meng Z~Y 2022 {\em
  Phys. Rev. B\/} {\bf 105} L041111

\bibitem{Anders2020}
Sandvik A~W and Zhao B 2020 {\em Chinese Physics Letters\/} {\bf 37} 057502

\end{thebibliography}

\end{document}